\documentstyle[]{l-aa} \input epsf 
\def\gsim{\lower.4ex\hbox{$\;\buildrel >\over{\scriptstyle\sim}\;$}}
\def\lsim{\lower.4ex\hbox{$\;\buildrel <\over{\scriptstyle\sim}\;$}}

  \def\bib{\bibitem{}}
\newcommand{\xia}{\overline{\xi}}
\newcommand{\rhoa}{\overline{\rho}}
\newcommand{\gam}{\gamma}

\newcommand{\law}{\stackrel{\mbox{law}}{=}}

\newcommand{\beq}{\begin{equation}}
\newcommand{\eeq}{\end{equation}}
\begin{document}
%
% US additions
% 
\topmargin=2.5 cm

\thesaurus{02(12.12.1; 11.03.1)} 
\title{Structure formation: a spherical model for the evolution of the 
density distribution}
\author{P. Valageas}
\institute{Service de Physique Th\'eorique, CEA Saclay, 91191
Gif-sur-Yvette, France \\ Theoretical Astrophysics Center, Juliane Maries Vej 30, 
2100 Copenhagen 0, Denmark}
\date{Received November 25, 1997; accepted May 12, 1998.}
\maketitle 
%\markboth{P. Valageas}{A spherical model for the evolution of the
%density distribution}

\begin{abstract}

Within the framework of hierarchical clustering we show that a simple
Press-Schechter-like approximation, based on spherical dynamics,
provides a good estimate of the evolution of the density field in the
quasi-linear regime up to $\Sigma \sim 1$. Moreover, it allows one to
recover the exact series of the cumulants of the probability
distribution of the density contrast in the limit $\Sigma \rightarrow
0$ which sheds some light on the rigorous result and on
``filtering''. We also obtain similar results for the divergence of
the velocity field.

Next, we extend this prescription to the highly non-linear regime,
using a stable-clustering approximation. Then we recover a specific
scaling of the counts-in-cells which is indeed seen in numerical
simulations, over a well-defined range. To this order we also
introduce an explicit treatment of the behaviour of underdensities,
which takes care of the normalization and is linked to the
low-density bubbles and the walls one can see in numerical simulations. We
compare this to a 1-dimensional adhesion model, and we present the consequences of our prescription for the power-law tail and the cutoff of the density
distribution.

\end{abstract}


\section{Introduction}

In the standard cosmological model gravitational structures in the universe
have formed by the growth of small density fluctuations present in the
early universe (Peebles 1980). These perturbations may be due to quantum
mechanical effects and are likely to be gaussian, so that they are described
by their power-spectrum. In many cases, like the CDM model (Peebles 1982;
Davis et al.1985), the power increases at small scales which leads to a
hierarchical scenario of structure formation. Small scales collapse first,
building small virialized objects which merge later to form broader and
broader halos as larger scales become non-linear. These halos will produce
galaxies or clusters of galaxies, according to the scale and other physical
constraints like cooling processes. Hence it is of great interest to
understand the evolution with time of the density field and the mass
functions of various astrophysical objects it implies. This is a necessary
step in order to model for instance the formation of galaxies or clusters,
which can later put constraints on the cosmological parameters using the
observed luminosity function of galaxies or the QSO absorption lines.

Within this hierarchical framework an analytical model for the mass
function of collapsed (or just-virialized) objects was proposed by Press \&
Schechter (1974) (hereafter PS). Numerical simulations (Efstathiou et
al.1988; Kauffmann \& White 1993) have shown this mass function is similar to the numerical results, although the agreement is improved by a modification of the usual density threshold $\delta_c$ used in this model (Lacey \& Cole 1994). However, this prescription encounters some serious defects, like
the cloud-in-cloud problem studied by Bond et al.(1991), and a well-known
normalization problem since one only counts half of the mass of the
universe. Moreover, underdense regions are not modelled by this approach,
and are simply taken into account in fine by a global multiplication of the
mass function by a factor 2 which allows to get the correct
normalization (this multiplicative factor 2 was recovered more rigorously by Bond et al.1991 for a top-hat in $k$, but this does not extend to more 
realistic filters). Finally, this model only predicts the mass function of
collapsed objects, while one may also be interested in mass condensations
of various density, in order to study galaxies or even underdense
structures (``voids'') for instance.

A method to describe the density field itself, and then derive the
multiplicity functions of any objects, is to consider the
counts-in-cells. These are closely related to the many-body correlation
functions and were studied in detail by Balian \& Schaeffer (1989) in the
highly non-linear regime within the framework of the stable clustering
picture (Peebles 1980), where the correlation functions are
scale-invariant. This assumption and the scaling it implies for the matter
distribution are indeed verified by numerical simulations (Bouchet et
al.1991; Colombi et al.1995) and observations (Maurogordato et
al.1992). Bernardeau \& Schaeffer (1991) and Valageas \& Schaeffer (1997)
(hereafter VS) described some of the consequences of this model for the
multiplicity functions of various objects like clusters or
galaxies. Indeed, a great advantage of this approach is that one can
study many different astrophysical objects (and even ``voids'') from a
unique model which is not restricted to just-collapsed objects. Moreover, it
is based on a rather general assumption, which does not depend on the exact
details of the dynamics, and VS showed that the PS prescription can be
recovered as a particular case among the possible models this
scale-invariant approach can describe. Note that since the PS approach is
unlikely to give a very precise description of the clustering process,
because of its simplicity and the problems it encounters, the
scale-invariant method offers the advantage to provide a simple and natural
way to take into account the possible corrections to the PS prescription.

In this article, we intend to show how a simple PS-like model, based on the
spherical dynamics, can describe the evolution of the density field and
provide a specific model for the scale-invariant picture studied in Balian
\& Schaeffer (1989) or VS. Thus, we consider both the quasi-linear ($\Sigma
< 1$) and highly non-linear ($\Sigma \gg 1$) regimes (where as usual $\Sigma(M,a)$ is the amplitude of the density fluctuations at scale $M$ given by the linear theory at the considered time), to relate the final
non-linear density field to the initial conditions and to show how the
scale-invariance of the many-body correlation functions and the scalings of
the multiplicity functions described in VS can arise from the hierarchical
structure formation scenario. We focus on the counts-in-cells, which
provide a powerfull description of the density field and can be used to
obtain any mass function of interest as detailed in VS. First, we derive the
statistics of the counts-in-cells which our spherical model implies in the
quasi-linear regime (Sect.2). We show in particular that we recover in the
limit $\Sigma \rightarrow 0$ the whole series of the cumulants derived
rigorously by Bernardeau (1994a) from the exact equations of motion. We also
consider the predictions of this model for the statistics of the divergence
of the velocity field (Sect.3) and explain why this simple spherical dynamics
works so well in the limit $\Sigma \rightarrow 0$ (Sect.4). Finally, we study
the non-linear regime (Sect.5). In addition to the virialization process of
overdensities we take care explicitely of the evolution of
underdensities. This new prescription solves in a natural way the
normalization problem (in the sense that all the mass will eventually get embedded within overdense halos) and is compared with a 1-dimensional adhesion
model. It also allows to complete the comparison with the scale-invariant
approach. Indeed, although we do not obtain as detailed results for
the mass function as the PS approach, which we believe anyway to be rather
illusory since such simple and crude descriptions cannot provide rigorous
and exact predictions, we think our model allows to understand  how scaling properties appear, and it can predict asymptotic behaviours like the slope 
of the power-law tail of the mass function or its exponential cutoff 
(so that the model can be tested). Moreover, it gives a usefull reference which would enable one to evaluate the magnitude of the effects neglected here from a comparison with the results obtained by a more rigorous calculation.

\section{Quasi-linear regime: density contrast}

As VS showed, it is possible to extend the usual PS approach to obtain an
approximation of the density field and the counts-in-cells it implies, in
addition to the mass function of just-virialized objects one generally
considers. Although VS focused on the highly non-linear regime (using the stable-clustering ansatz), we shall
here first consider the simpler case of the quasi-linear regime $\Sigma < 1$.

\subsection{Critical universe: $\Omega=1$}

\subsubsection{Lagrangian point of view}

We first consider a Lagrangian point of view, which is well suited to
PS-like approaches since the fundamental hypothesis of such approximations
is that it is possible to follow the evolution of fluid elements
recognized in the early linear universe (one usually only considers
overdensities) up to the non-linear regime. Thus, the usual PS
prescription assumes i) that the dynamics of these objects is given by
the spherical model and ii) that their keep their identity throughout
their evolution. Hence, to any piece of matter identified in the early
linear universe one can assign at a later time a specified radius and
density. Of course, this cannot be exact as all particles cannot
follow simultaneously a spherical dynamics, and this approach cannot
take into account mergings which change the number of objects and
destroy their identity. However, we shall see below that it can
provide a reasonable estimate of the early density field.

In this Lagrangian approach, we consider the evolution of ``objects'' of
mass $M$, identified in the early linear universe. In other words, the
filtering scale is the mass and not the radius. According to the spherical
model, particles which were initially embedded in a spherical region of
space characterized by the density contrast $\delta_L$ ($\delta_L$ is the
density contrast given by the linear theory at any considered time:
$\delta_L \propto a$ where $a(t)$ is the scale factor) find themselves in a
spherical region of density contrast $\delta$ at the same scale $M$, given
at any epoch by: 
\beq 
\delta = {\cal F}(\delta_L) 
\eeq 
The function ${\cal F}$ is defined by the dynamics of the spherical model
(Peebles 1980): 
\beq 
\delta_L > 0 \; : \;\; \left\{ \begin{array}{rl}
1+\delta & = {\displaystyle \frac{9}{2} \; \frac{(\theta-\sin
\theta)^2}{(1-\cos \theta)^3} } \\ \\ \delta_L & = {\displaystyle
\frac{3}{20} \; \left[ 6 (\theta-\sin \theta) \right]^{2/3} } \end{array}
\right.  
\eeq 
and 
\beq 
\delta_L < 0 \; : \;\; \left\{ \begin{array}{rl}
1+\delta & = {\displaystyle \frac{9}{2} \; \frac{(\sinh
\eta-\eta)^2}{(\cosh \eta-1)^3} } \\ \\ \delta_L & = {\displaystyle -
\frac{3}{20} \; \left[ 6 (\sinh \eta-\eta) \right]^{2/3} } \end{array}
\right.  \label{undersph} 
\eeq 
This definition of ${\cal F}(\delta_L)$ breaks down for $\delta_L \geq
\delta_c$, with $\delta_c = 3/20 \; (12 \pi)^{2/3} \simeq 1.69$, where
$\delta$ becomes infinite. This is usually cured by a virialization
prescription: the halo is assumed to virialize in a finite radius taken for
instance as one half its radius of maximum expansion, so that ${\cal
F}(\delta_c) = \Delta_c$ with $1+\Delta_c=18 \pi^2 \simeq 178$. However, we
shall not consider this modification in this section, as we focus here on
the quasi-linear regime where most of the mass is embedded within regions
of space with a density contrast $\delta$ smaller than $\Delta_c$, hence we
restrict ourselves to $\delta < \Delta_c$. Now, we define the Lagrangian
probability distribution $P_m(\delta)$: if we choose at random a spherical 
region of mass $M$ its density contrast is between $\delta$ and 
$\delta+d\delta$ with probability $P_m(\delta) \; d\delta$ (here the index $m$ does not stand for a given mass scale, its use is only to distinguish the Lagrangian quantities used here, where we follow matter elements, from their Eulerian counterparts which we shall introduce in the next sections). 
Within the framework of the spherical model, this probability distribution is simply given by: 
\beq 
P_m(\delta) \; d\delta =
P_L(\delta_L) \; d\delta_L \hspace{0.5cm} \mbox{with} \hspace{0.5cm} \delta
= {\cal F}(\delta_L) \label{PM1} 
\eeq 
where $P_L(\delta_L)$ is the probability distribution relative to the
linear density contrast, or the early universe. We shall assume that the
initial density fluctuations are gaussian, so that: 
\beq 
P_L(\delta_L) =
\frac{1}{\sqrt{2\pi}\Sigma} \; e^{-\delta_L^2/(2\Sigma^2)} \label{Pgaus}
\eeq 
where as usual $\Sigma=\Sigma(M,a)$ is the amplitude of the density
fluctuations at scale $M$ given by the linear theory at the considered
time ($\Sigma \propto a \; M^{-(n+3)/6}$ for an initial power-spectrum
which is a power-law: $P(k) \propto k^n$). We can note that the
probability distribution $P_m(\delta)$ is correctly normalized: $\int
P_m(\delta) \; d\delta = 1$ by definition. However, the mean value of
$\delta$ is not equal to 0. This is quite natural since particles get
embedded within increasingly high overdensities (while the density
contrast cannot be smaller than $-1$), and at late times one expects
that all the mass will be part of overdense virialized halos. However,
this late stage is not described by the model introduced in this section, 
which does not
include virialization and where half of the mass remains at any time
within underdensities (this is the well-known normalization problem of
the PS prescription by a factor 2).

We can also characterize the probability distribution $P_m(\delta)$ by the
generating function $\varphi_m(y,\xia_m)$ (and by $\xia_m$) which we define by: 
\beq
e^{-\varphi_m(y,\xia_m)/\xia_m} = \int_{-1}^{\infty} e^{-\delta \;
y/\xia_m} \;\; P_m(\delta) \; d\delta \label{phiMPM} 
\eeq 
where $\xia_m = < \delta^2 >$. The function $\varphi_m(y,\xia_m)$ generates
the series of the moments of $P_m(\delta)$: 
\beq
e^{-\varphi_m(y,\xia_m)/\xia_m} = 1 + \sum_{p=1}^{\infty} \frac{(-1)^p}{p!}
\; y^p \; \frac{<\delta^p>}{\xia_m^{\;p}} 
\eeq 
In the highly non-linear regime considered by VS the function
$\varphi_m(y,\xia_m)$ was assumed to be scale-invariant, so that it did not
depend on $\xia_m$, but here this is not the case. If $<\delta>=0$ the
usual series of the cumulants is generated by $\varphi(y,\xia_m)$: 
\beq
<\delta>=0 \; : \;\; \varphi_m(y,\xia_m) = \sum_{p=2}^{\infty}
\frac{(-1)^{p-1}}{p!} \; y^p \; \frac{<\delta^p>_c}{\xia_m^{\;p-1}} 
\eeq
The relation (\ref{phiMPM}) can be written in terms of the linear density
contrast $\delta_L$:
\[
e^{-\varphi_m(y,\xia_m)/\xia_m} = \int_{-\infty}^{\infty}
\frac{d\delta_L}{\sqrt{2\pi} \Sigma}
\]
\beq
\hspace{2cm} \times \; \exp \left[ - \frac{1}{\Sigma^2} \left(
\frac{\delta_L^2}{2} + {\cal F}(\delta_L) \; y \; \frac{\Sigma^2}{\xia_m}
\right) \right]   \label{phimSig}
\eeq 
In the quasi-linear regime, that is for $\xia_m \rightarrow 0$, $\Sigma
\rightarrow 0$ and $\Sigma^2/\xia_m \rightarrow 1$, we can use the saddle
point method to get: 
\beq 
\left\{ \begin{array}{rl} \varphi_m(y) & = {\displaystyle y \; {\cal
F}(\delta_y) \; + \; \delta_y^2/2 } \\ \\ \delta_y & = {\displaystyle - y
\; {\cal F}'(\delta_y) } \end{array} \right.  \label{phiM} 
\eeq 
where $\varphi_m(y)$ is the limit of $\varphi_m(y,\xia_m)$ for $\xia_m
\rightarrow 0$. If we define $\tau_m = -\delta_y$ and ${\cal G}_m(\tau_m) =
{\cal F}(\delta_y)$, we obtain 
\beq 
\left\{ \begin{array}{rl} \varphi_m(y) & = {\displaystyle y \; {\cal
G}_m(\tau_m) \; - \; y \; \tau_m \; {\cal G}'_m(\tau_m) / 2 } \\ \\ \tau_m
& = {\displaystyle - y \; {\cal G}'_m(\tau_m) } \end{array} \right.  
\eeq
This is exactly the result derived rigorously by Bernardeau (1994a) from the equations of motion, when the matter is described by a pressure-less
fluid. Hence our approach should give a good description of the early
stages of gravitational clustering, since it is leads to the right limit
for the cumulants in the linear regime, which was not obvious at first
sight. Moreover, it provides some hindsight into the result of the exact
calculation, as the function ${\cal F}$ or ${\cal G}_m$ which represents
the spherical dynamics appears naturally in the expression of
$\varphi_m(y)$. We shall come back to this point in Sect.4. We can note that
our model implies in addition a specific dependence of
$\varphi_m(y,\xia_m)$ on $\xia_m$ (or $\Sigma$), which could be computed from (\ref{phimSig}).

\subsubsection{Eulerian point of view}

For practical purposes, one is in fact more interested in the Eulerian
properties of the density field, where the filtering scale is a length
scale $R$. For instance, a convenient way to describe the density
fluctuations is to consider the counts-in-cells: one divides the universe
into cells of scale $R$ (and volume $V$) and defines the probability
distribution $P(\delta)$ of the density contrast within these cells. Hence,
we need to relate this Eulerian description to the Lagrangian model we
developed in the previous section. As was done in VS, we shall use: 
\beq 
\int_{\delta}^{\infty} \; (1+\delta') \; P(\delta') \; d\delta' =
F_m(>\delta,M) \label{EuLa} 
\eeq 
with $M = (1+\delta) \; \rhoa \; V$. Thus the mass
embedded in cells of scale $R$ with a density contrast larger than
$\delta$ is taken equal to the mass formed by particles
which are located within spherical regions of scale $M$
with a density contrast larger than $\delta$. Using our Lagrangian model, 
we get:
\beq 
F_m(>\delta,M) = F_{mL}(>\delta_L,M) \;\;\; \mbox{with} \;\;\;
\delta = {\cal F}(\delta_L)  \label{FmFmLM}
\eeq 
where $F_{mL}$ is the mass fraction obtained in the linearly
extrapolated universe. From (\ref{EuLa}) and (\ref{Pgaus}) we have:
\beq 
(1+\delta) \; P(\delta) = - \; \frac{\partial}{\partial \delta} \;
\int_{\delta_L/\Sigma(R_m)}^{\infty} e^{-\nu^2/2} \;
\frac{d\nu}{\sqrt{2\pi}} \label{Pnu1} 
\eeq 
with: 
\beq 
\left\{
\begin{array}{rcl} \delta & = & {\cal F}[\delta_L] \\ \\ R_m^3 & = &
(1+\delta) \; R^3 \end{array} \right.  \label{deltLdelRmR}
\eeq 
Finally, we obtain: 
\beq
P(\delta) = \frac{1}{\sqrt{2\pi}} \; \frac{1}{1+\delta} \;
\frac{\partial}{\partial \delta} \left[ \frac{\delta_L}{\Sigma(R_m)}
\right] \; \exp\left[-\frac{1}{2} \left( \frac{\delta_L}{\Sigma}
\right)^2 \right] \label{Pdel} 
\eeq 
as in VS. Of course we recover all the mass of the universe $\int
(1+\delta) \; P(\delta) \; d\delta = 1$. In fact, half of the mass is
in overdensities and the other half in underdensities, as was the case
in the Lagrangian description. However, in general this probability
distribution is not correctly normalized: $\int P(\delta) \; d\delta
\neq 1$ (but in the linear regime, $\xia \rightarrow 0$ and $\Sigma
\rightarrow 0$, its normalization tends to unity).

We can still define a generating function $\varphi(y,\xia)$, as in
(\ref{phiMPM}), which leads to:
\beq
\begin{array}{ll} {\displaystyle
e^{-\varphi(y,\xia)/\xia} } & = {\displaystyle  \int_{-1}^{\infty}
\frac{d\delta}{\sqrt{2\pi}} \; \frac{1}{1+\delta} \;
\frac{\partial}{\partial \delta} \left[ \frac{\delta_L}{\Sigma(R_m)}
\right] } \\  \\ & {\displaystyle 
\hspace{0.5cm} \times \; \exp \left[ - \frac{1}{\xia} \left( \delta \;
y \; + \; \frac{\delta_L^2}{2} \; \frac{\xia}{\Sigma^2} \right)
\right]  } \end{array}    \label{phieul} 
\eeq 
Using $\delta = {\cal F}(\delta_L) = {\cal G}_m(\tau_m)$, with $\tau_m
= -\delta_L$, we define: 
\beq 
\left\{
\begin{array}{rcl} {\cal G}(\tau) & = & {\displaystyle {\cal G}_m
\left( \tau \; \frac{ \Sigma \left[ (1+{\cal G}(\tau))^{1/3} \; R
\right] } {\sqrt{\xia}} \right) } \\ \\ \tau(\delta) & = &
{\displaystyle \frac{\tau_m \; \sqrt{\xia}}{\Sigma \left[ (1+{\cal
G}(\tau))^{1/3} \; R \right] } } \end{array} \right.  
\eeq 
Hence ${\cal G}_m(\tau_m) = {\cal G}(\tau) = \delta$. In the
quasi-linear regime, that is for $\xia \rightarrow 0$, $\Sigma
\rightarrow 0$ and $\Sigma^2/\xia \rightarrow 1$, the saddle point
method leads to: 
\beq 
\left\{ \begin{array}{rl} \varphi(y) & = {\displaystyle y \; {\cal
G}(\tau) \; - \; y \; \tau \; {\cal G}'(\tau) / 2 } \\ \\ \tau & =
{\displaystyle - y \; {\cal G}'(\tau) }
\end{array} \right.  
\eeq 
where $\varphi(y)$ is the limit of $\varphi(y,\xia)$ for $\xia
\rightarrow 0$. Thus, once again we recover the result derived
rigorously by Bernardeau (1994a). Note that we should have modified
$P(\delta)$ in (\ref{phieul}) since it is not correctly normalized,
however if this modification only consists of a multiplication factor
(which must go to unity in the limit $\xia \rightarrow 0$) and a
change of the shape of $P(\delta)$ for density contrasts much larger
than $\xia$ as $\xia \rightarrow 0$, this does not modify the function
$\varphi(y)$ we obtained. If the power-spectrum is a power-law $P(k)
\propto k^n$, (\ref{Pdel}) can be written:
\[
P(\delta) = \frac{(1+\delta)^{(n-3)/6}}{\sqrt{2\pi} \Sigma(R)} \left[
\frac{1}{{\cal F}'(\delta_L)} + \frac{3+n}{6}
\frac{\delta_L}{1+\delta} \right]
\] 
\beq
\hspace{2cm} \times \; \exp \left[ - \frac{1}{2} \left(
\frac{\delta_L}{\Sigma(R)} (1+\delta)^{(3+n)/6} \right)^2 \right]
\label{Pd2Om1} 
\eeq 
If the power-spectrum is not a power-law, our method is still valid
and we have to use (\ref{Pdel}). However, for a very smooth $P(k)$,
like a CDM power-spectrum, an easier and still reasonable
approximation is to use (\ref{Pd2Om1}) where $n$ is the local slope of
the power-spectrum at scale $R$. We shall compare this approximation
to numerical results in Sect.2.3. The fact that we recover the exact 
generating function
$\varphi(y,\xia)$ for $\xia \rightarrow 0$ suggests again that our
prescription could provide reasonable results in the early linear
universe when $\xia < 1$. We shall see below that it is indeed the
case, by a comparison with numerical simulations. We can note that our
probability distributions (\ref{Pd2Om1}) look rather different from those
obtained by Bernardeau (1994a) since the high-density cutoff is usually
different from a simple exponential. However, they both agree with numerical
results for $\delta \ll \Delta_c$. In fact, neither of these approaches 
should be used for large density contrasts
$\delta > \Delta_c$ where shell-crossing and virialization play an
important role.

\subsection{Open universe: $\Omega<1, \; \Lambda=0$}

\subsubsection{Lagrangian point of view}

	In the case of an open universe, we can still apply the method
described previously for a critical universe but there is now an
additional time dependence in the relation $\delta_L - \delta$. Thus,
(\ref{PM1}) becomes: 
\beq 
P_m(\delta) \; d\delta = P_L(\delta_L) \; d\delta_L \hspace{0.8cm} \mbox{with}
\hspace{0.8cm} \delta = {\cal F}(\delta_L,a)
\eeq
In the specific case where $\Lambda=0$, one defines: \beq \eta_b =
\cosh^{-1} \left( \frac{2}{\Omega} -1 \right) \eeq and \beq D(t) =
\frac{3 \sinh \eta_b (\sinh \eta_b - \eta_b)}{(\cosh \eta_b-1)^2} -2
\eeq which is the growing mode of the linear approximation normalized
so that $D(t \rightarrow \infty)=1$. Then, the function ${\cal
F}(\delta_L,a)$ is given by:
\[
\delta_L > \frac{3}{2} D(t) \left\{ \begin{array}{rl} 1+\delta & =
{\displaystyle \left( \frac{\cosh \eta_b -1}{1-\cos \theta} \right)^3
\; \left( \frac{\theta-\sin \theta}{\sinh \eta_b-\eta_b} \right)^2 }
\\ \\ \delta_L & = {\displaystyle \frac{3}{2} \; D(t) \; \left[ 1 +
\left( \frac{\theta-\sin \theta}{\sinh \eta_b - \eta_b} \right)^{2/3}
\right] } \end{array} \right.
\]
and
\[
\delta_L < \frac{3}{2} D(t) \left\{ \begin{array}{rl} 1+\delta & =
{\displaystyle \left( \frac{\cosh \eta_b -1}{\cosh \eta-1} \right)^3
\; \left( \frac{\sinh \eta- \eta}{\sinh \eta_b-\eta_b} \right)^2 } \\
\\ \delta_L & = {\displaystyle \frac{3}{2} \; D(t) \; \left[ 1 -
\left( \frac{\sinh \eta- \eta}{\sinh \eta_b - \eta_b} \right)^{2/3}
\right] } \end{array} \right.
\]

In the case $\Omega \rightarrow 0$ ($\eta_b \rightarrow \infty, \;
D(t) \rightarrow 1$), large overdensities have already collapsed, and
we are left with: 
\beq 
\Omega = 0 \; : \;\; 1+\delta = \left( 1 - \frac{2}{3} \delta_L
\right)^{-3/2} \label{F0} 
\eeq 
This simple form for ${\cal F}(\delta_L)$ leads to: 
\beq 
P_m(\delta) = \frac{(1+\delta)^{-5/3}}{\sqrt{2\pi} \Sigma(M)} \; \exp
\left[ - \frac{9}{8 \Sigma^2} \left( 1 - (1+\delta)^{-2/3} \right)^2
\right] 
\eeq 
which provides a convenient estimation for $P_m(\delta)$. As for a
critical universe we can still define a generating function
$\varphi_m(y,\xia_m)$ which allows us to recover the results of
Bernardeau (1994a).

\subsubsection{Eulerian point of view}

Naturally, we can obtain an approximation for the probability
distribution $P(\delta)$ of the density contrast within cells of scale
$R$ in a fashion similar to what we did for a critical universe from
the Lagrangian probability $P_m(\delta)$. In the case of a
power-spectrum which is a power-law, and in the limit $\Omega
\rightarrow 0$, we get for instance:
\beq
\begin{array}{ll} {\displaystyle 
P(\delta)} & {\displaystyle = \frac{(1+\delta)^{(n-9)/6}}{4 \sqrt{2\pi} \Sigma(R)} \;
\left[ n+3 + (1-n) (1+\delta)^{-2/3} \right] }
\\ \\
 & {\displaystyle \times \; \exp \left[ - \frac{9 (1+\delta)^{(n+3)/3}}{8
\Sigma^2} \left( 1 - (1+\delta)^{-2/3} \right)^2 \right] } \end{array} \label{PdOm0}
\eeq 
As we shall see in the next section, this very simple formula provides
in fact a good approximation to $P(\delta)$ for all values of $\Omega$
of interest (even for $\Omega=1$) due to the weak dependence of ${\cal
F}(\delta_L,\Omega)$ on $\Omega$. As for a critical universe we can
also define a generating function $\varphi(y,\xia)$ and recover the
results of Bernardeau (1994a).

We can note that Protogeros \& Scherrer (1997) obtained similar
results with an approach close to ours. They considered ``local
Lagrangian approximations'' where the density contrast at the
Lagrangian point ${\bf q}$, time $t$, is related to its initial value
by $\delta = {\cal F}(\delta_L)$. They used several approximations for
${\cal F}$ including the simplified spherical collapse model
(\ref{F0}) and obtained the probability distribution (\ref{PdOm0}),
which they modified by introducing an ad-hoc multiplicative function $N(t)$ 
within the relation $\delta_L \rightarrow (1+\delta)$ in order
to normalize properly $P(\delta)$. However, our model differs from
theirs by some aspects. Thus, our Lagrangian probability distribution
is defined from the start with respect to a given mass scale $M$
(which might be seen as a ``smoothing'' scale). This appears naturally
in our approach, and it ensures we always work with well-defined
quantities. Indeed, for a power-spectrum which is a pure power-law
with $n>-3$ the ``unsmoothed'' density field is not a function but a
distribution. In fact, one cannot characterize a
point by a finite density, without specifying the scale over which
this density is realised. Then, the
change from the Lagrangian to the Eulerian view-point is quite
natural, and it provides an Eulerian distribution function which
differs from the Lagrangian one in a very simple and physical manner
and which depends on the power-spectrum. No smoothing procedure needs to be
applied in fine in order to compare with observations: a filtering scale
($M$ or $R$) is always automatically included in our approach. 
Our approximation also shows
clearly the dependence on $\Omega$ of the functions ${\cal
F}(\delta_L)$ and $\varphi(y)$. Finally, one can note that contrary to
Protogeros \& Scherrer (1997) we did not normalize our probability
distribution $P(\delta)$. In fact, we think such a procedure is somewhat
artificial and may lead to an even worse approximation. Indeed, if the
normalization problem comes mainly from a specific interval of the
density contrast where our approximation is very bad, a simple
normalization procedure will not give very accurate results in this
interval (since the starting values have no relation with the correct
ones) while it will destroy our predictions in the interval where they
were fine. Thus, one can fear such ``cure'' may in fact spread errors
over all density contrasts. We shall come back to this point below,
but we can already note that for $n=1$ the probability distribution
(\ref{PdOm0}) cannot be meaningfully normalized since $\int P(\delta)
d\delta = \infty$.

\subsection{Comparison with numerical results}

\begin{figure}[htb]

\begin{picture}(230,170)(-18,-19) 

\epsfxsize=8.22 cm
\epsfysize=12 cm
\put(-14.5,-115){\epsfbox{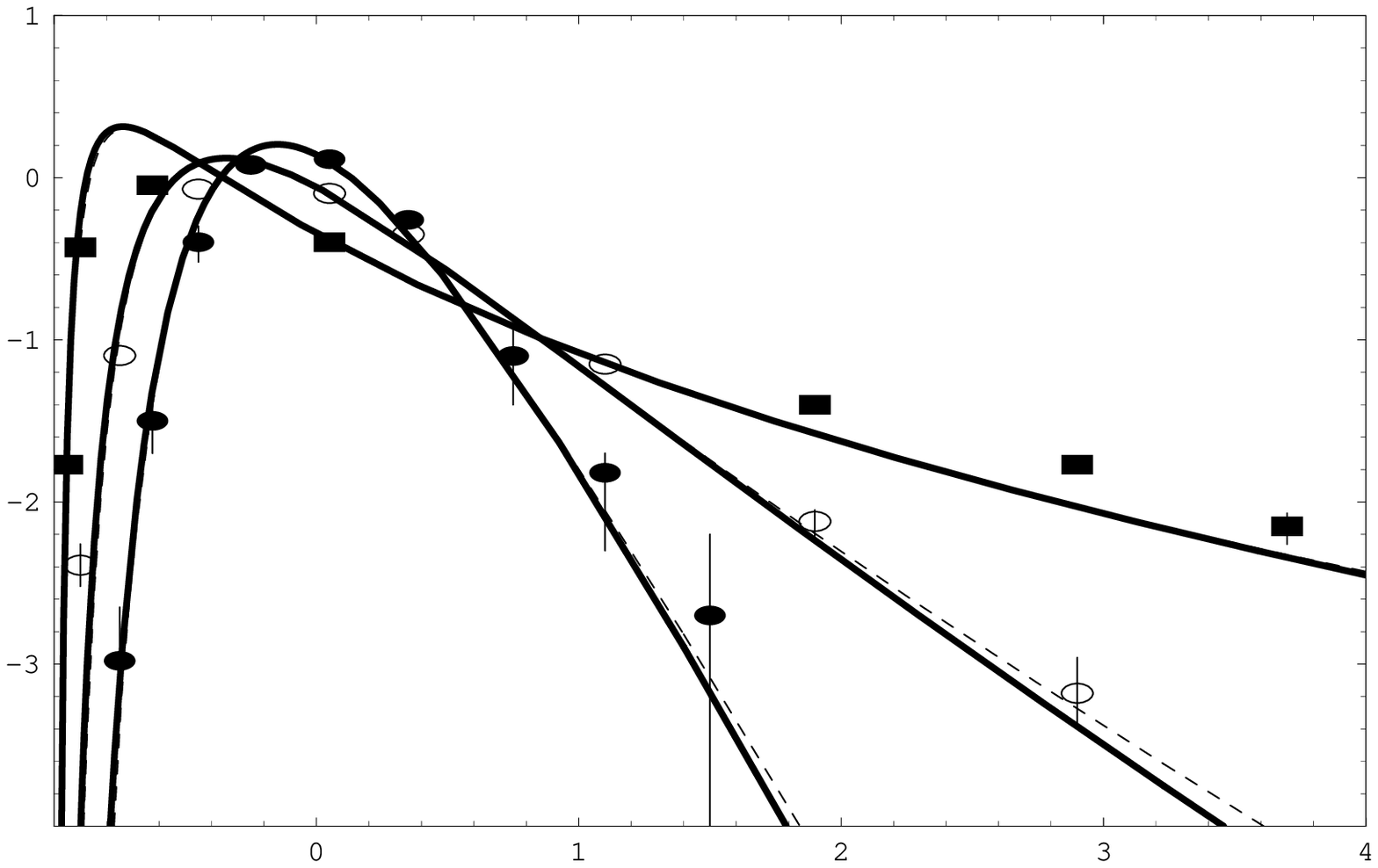}} 
\put(89,140){$\log[P(\delta)]$}
\put(101,-31){$\delta$}
\put(137,70){$\sqrt{\xia} = 0.92$}
\put(140,60){$n=-1.3$}
\put(171,15){$\sqrt{\xia} = 0.46$}
\put(182,5){$n=-1$}
\put(55,5){$\sqrt{\xia} = 0.29$}
\put(60,-5){$n=-0.7$}

\end{picture}

\caption{The probability distribution of the density contrast
$P(\delta)$. The solid lines present the prediction of our
prescription for a critical universe and a power-spectrum which is a
power-law (relation (\ref{Pdel}) or (\ref{Pd2Om1})), for various
$\xia$ and $n$, while the dashed-lines show the corresponding curves
for an empty universe $\Omega=0$ (relation (\ref{PdOm0})). The data
points are taken from Bernardeau (1994a) and Bernardeau \& Kofman
(1995) and correspond to a numerical simulation with a CDM initial
power-spectrum in a critical universe (thus $n$ is the local slope of
the power-spectrum). The density fluctuation $\xia$ and $n$ were
measured in the simulation.}
\label{figPd1}

\end{figure}

\begin{figure}[htb]

\begin{picture}(230,170)(-18,-19)

\epsfxsize=8.22 cm
\epsfysize=12 cm
\put(-14.5,-115){\epsfbox{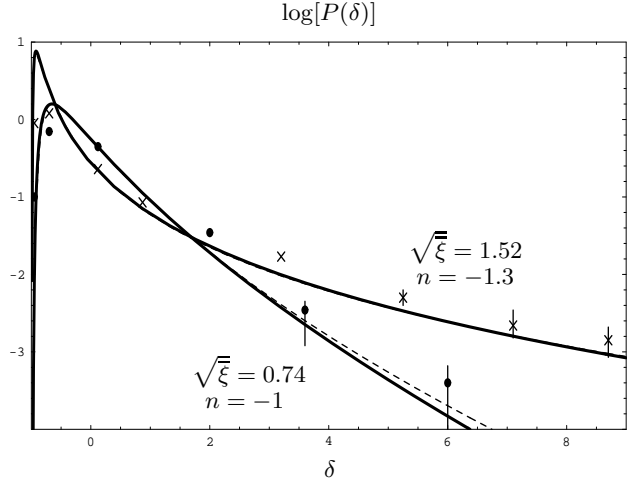}} 
\put(89,140){$\log[P(\delta)]$}
\put(105,-33){$\delta$}
\put(137,50){$\sqrt{\xia} = 1.52$}
\put(140,40){$n=-1.3$}
\put(55,3){$\sqrt{\xia} = 0.74$}
\put(60,-7){$n=-1$}

\end{picture}

\caption{The probability distribution of the density contrast
$P(\delta)$ as in Fig.\ref{figPd1} but for two different values of
$\xia$.}
\label{figPd2}

\end{figure}

From the results of previous paragraphs, since the functions
$\varphi(y)$ obtained with our prescription in the limit $\xia
\rightarrow 0$ are exactly those derived by a rigorous calculation we
can expect that the probability distribution $P(\delta)$ we get should
provide a good approximation in the linear regime. Thus,
Fig.\ref{figPd1} and Fig.\ref{figPd2} present a comparison of our
approximation with the results of numerical simulations, taken from
Bernardeau (1994a) and Bernardeau \& Kofman (1995) in the case of a CDM
initial power-spectrum in a critical universe. We display our
predictions for a critical universe (relation (\ref{Pd2Om1})) and an
``empty'' universe (relation (\ref{PdOm0})), for a power-spectrum
which is a power-law with $n$ given by the local slope of the actual
power-spectrum.

We can see that our approach leads indeed to satisfactory results up
to $\xia \sim 1$ given its extreme simplicity. As was noticed by
Bernardeau (1992), the $\Omega$ - dependence of the probability
distribution is very weak (the dashed lines are very close to the
solid lines on the figures), which means that the simple expression
(\ref{PdOm0}) provides a reasonable fit for all cosmological models up
to $\xia \sim 1$. However, as we can see on Fig.\ref{figPd2}, our
prescription leads to a sharp peak for very underdense regions $\delta
\simeq -1$ which increases with $\xia$ but does not appear in the
numerical results. This defect is also related to the fact that our
probability distribution is not correctly normalized to unity. This
problem is due to the expansion of underdense regions, which according
to the spherical dynamics grow faster than the average expansion of
the universe so that in our present model these areas occupy after
some time a volume which is larger than the total volume which is
available, which leads to a probability which is too large. Indeed,
within our approach underdense regions can expand without any limit
while in reality this growth is constrained by the fact that on large
scales we must recover the average expansion $\propto a^3$. Thus,
underdensities join together after some time and their mutual
influence alters their dynamics, which we did not take into account.

\begin{figure}[htb]

\begin{picture}(230,421)(0,-15)

\epsfxsize=8 cm
\epsfysize=8 cm
\put(12,230){\epsfbox{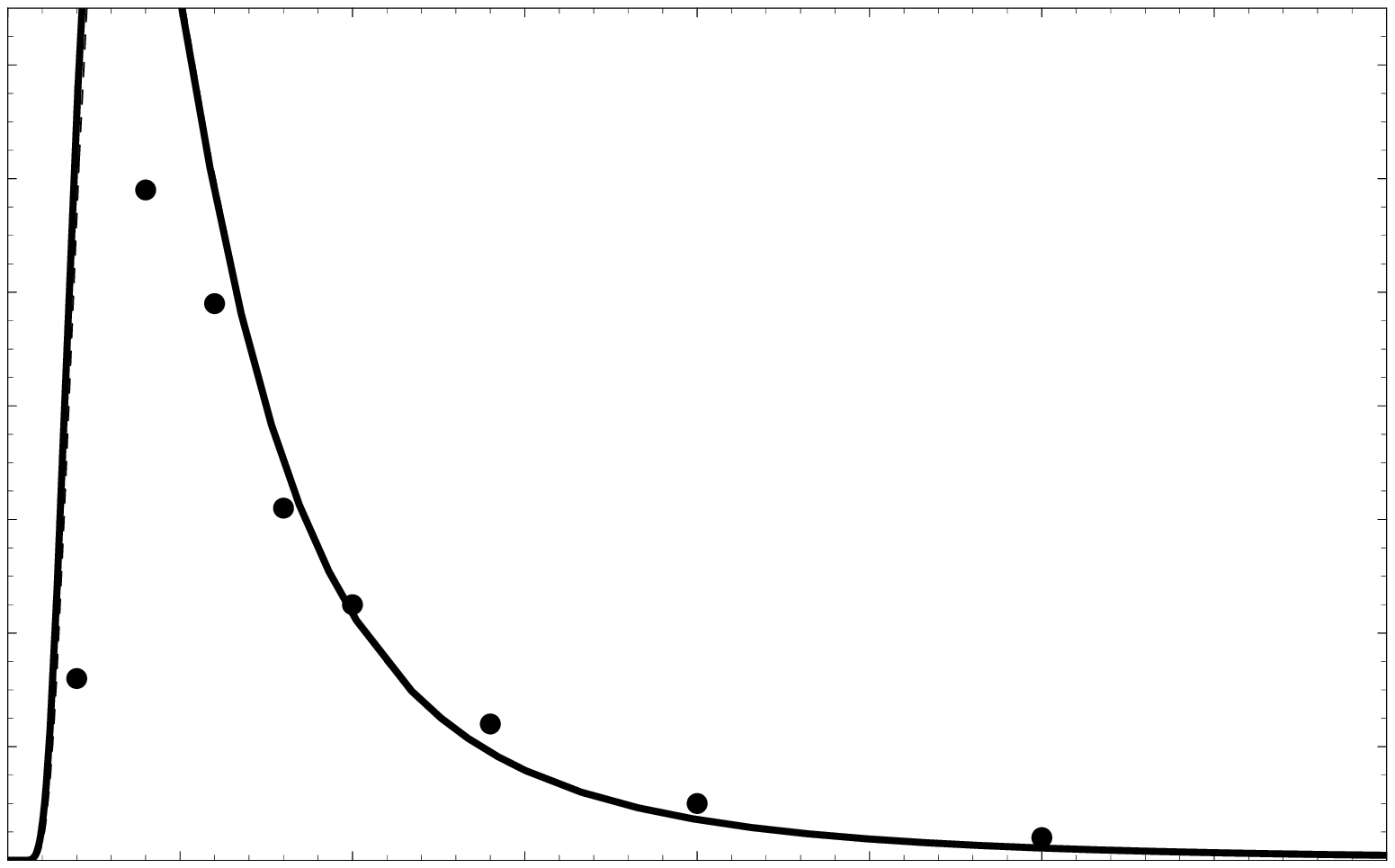}}
\put(13,295){0}
\put(7,334){0.6}
\put(7,372){1.2}
\put(140,370){$n=-2$}
\put(140,350){$\Sigma=0.92$}

\epsfxsize=8 cm
\epsfysize=8 cm
\put(12,130){\epsfbox{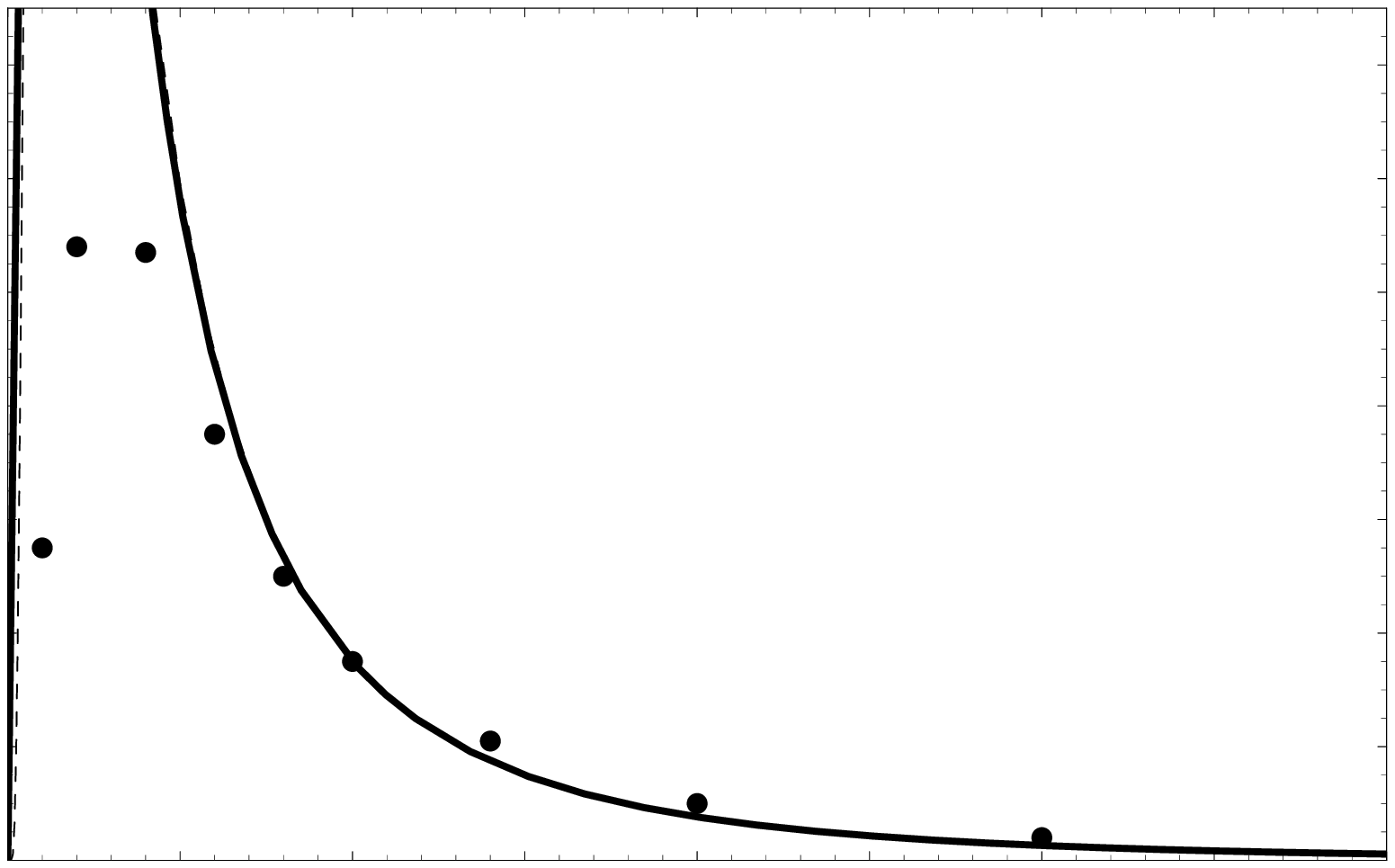}}
\put(13,195){0}
\put(7,234){0.6}
\put(7,272){1.2}
\put(140,270){$n=-1$}
\put(140,250){$\Sigma=1.14$}

\epsfxsize=8 cm
\epsfysize=8 cm
\put(12,30){\epsfbox{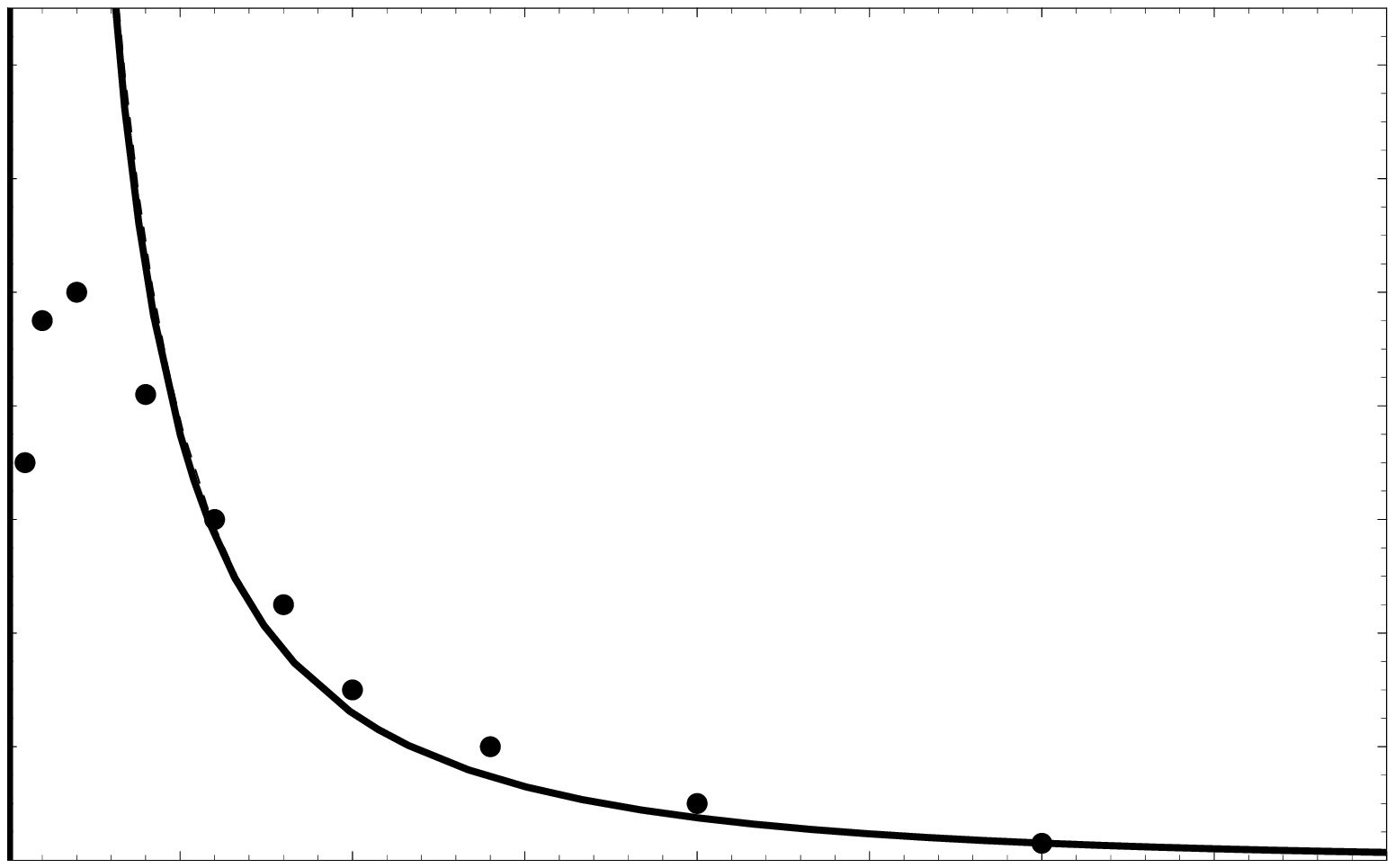}}
\put(13,95){0}
\put(7,134){0.6}
\put(7,172){1.2}
\put(140,170){$n=0$}
\put(140,150){$\Sigma=1.54$}

\epsfxsize=8 cm
\epsfysize=8 cm
\put(12,-70){\epsfbox{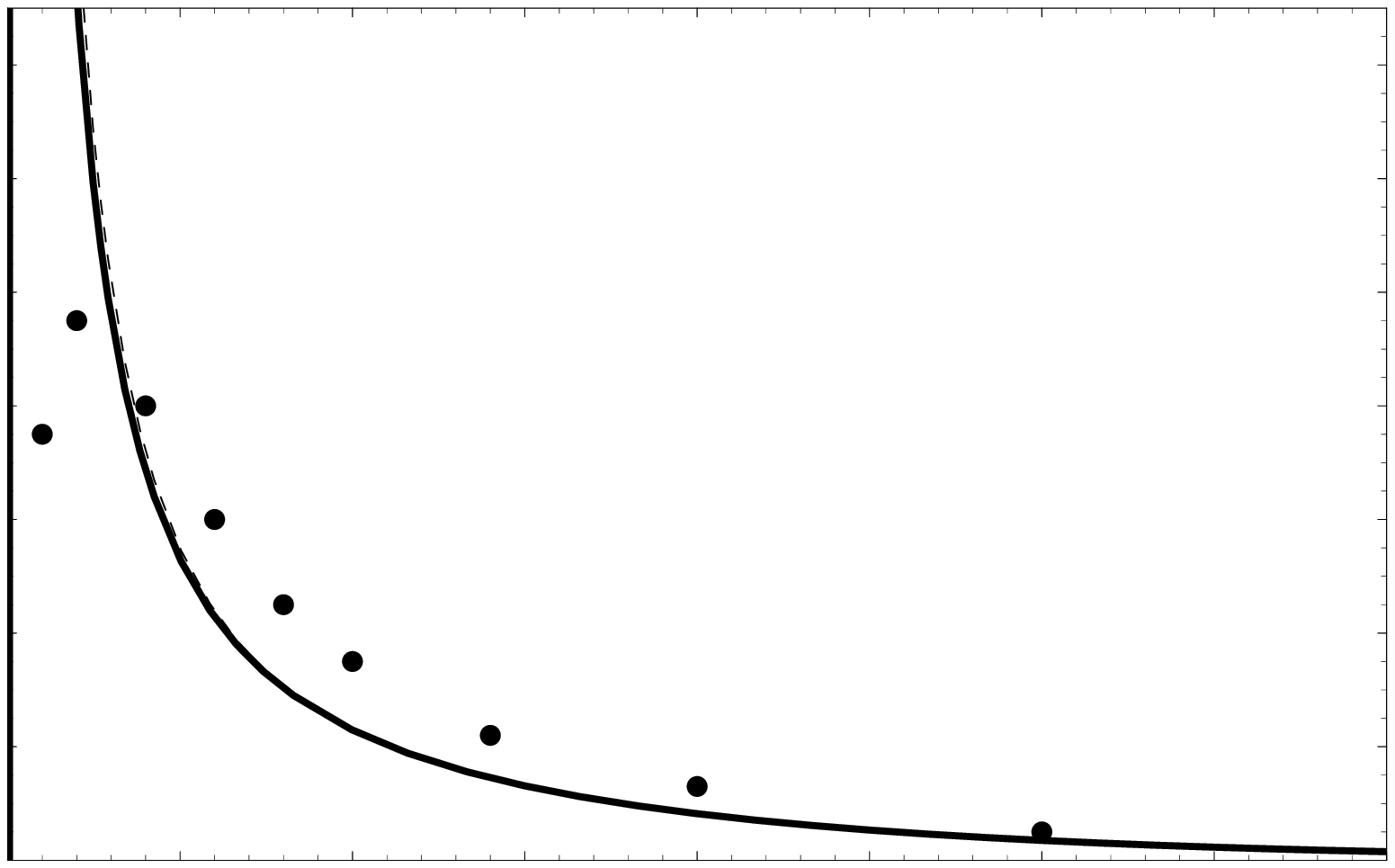}}
\put(75,-12){0}
\put(129,-12){1}
\put(182,-12){2}
\put(13,-5){0}
\put(7,34){0.6}
\put(7,72){1.2}
\put(140,70){$n=1$}
\put(140,50){$\Sigma=1.74$}
\put(120,403){$P(\delta)$}
\put(129,-25){$\delta$}

\end{picture}

\caption{The probability distribution of the density contrast
$P(\delta)$. The solid lines present the prediction of our
prescription for a critical universe and a power-spectrum which is a
power-law (relation (\ref{Pdel}) or (\ref{Pd2Om1})), for various
$\Sigma$ and $n$, while the dashed-lines show the corresponding curves
for an empty universe $\Omega=0$ (relation (\ref{PdOm0})). The data
points are taken from Protogeros et al.(1997).}
\label{figPMel}

\end{figure}

A simple way to normalize correctly the probability distribution
$P(\delta)$ would be to define the latter from the generating function
$\varphi(y,\xia)$ which one would take identical to $\varphi(y)$ for
any $\xia$: this is the method used successfully by Bernardeau
(1994a). However, there is a priori no fundamental reason why this
should be a particularly efficient procedure (except from the mere
constatation that it works). In fact, as we can see on
Fig.\ref{figPd2} we can expect most of the problem to come from the
peak which develops at low density contrasts, so that one should keep
the consequences of the evolution of $P(\delta)$, and
$\varphi(y,\xia)$, in other ranges of $\delta$ and simply disregard
the predictions obtained in the vicinity of this peak or introduce a
specific modification for this interval. This is even clearer on
Fig.\ref{figPMel} where we can see that our approximation can still
provide reasonable predictions for $\delta>0$ and $\Sigma \simeq 1$
although it completely fails for underdense regions. Of course the
agreement with numerical results improves for smaller $\Sigma$, as shown on Fig.\ref{figPd1} and Fig.\ref{figPd2}, and becomes excellent in the limit
$\Sigma \rightarrow 0$. We only show on Fig.\ref{figPMel} the 
largest values of $\Sigma$ where our approximation still makes some sense, in order to present its range of validity and to show clearly for which values
of $\delta$ (underdensities) it breaks down first. As
Protogeros et al.(1997) noticed, the problem becomes increasingly
severe as $n$ gets larger. However, our predictions work
better than those used by these authors because we did not normalize
our probability distribution (see their Fig.2). This was expected
since we noticed earlier that $\int P(\delta) d\delta = \infty$ for
$n=1$, so that it cannot even be normalized, but this does not
prevent our approximation to provide very good results for low
$\Sigma$, and when $\Sigma \rightarrow 0$ we recover the gaussian on
any finite interval of the density contrast which does not include
$\delta=-1$.

\section{Quasi-linear regime: velocity divergence}

The prescription we described in the previous paragraphs can also
predict the statistical properties of the divergence of the velocity
field. Note that we cannot get any reliable result for the shear,
since our model is based on a pure spherical dynamics, so that the
shear is zero for all the individual regions of matter we
consider. This is clearly an important shortcoming of this simple
approximation, however in the linear regime where density fluctuations
are small the rotational part of the velocity field decays so that
after a long time (but still in the linear regime) keeping only the
growing mode the velocity can be described by its divergence, or a
velocity potential. Thus, our model does not contradict a priori the
properties of the linear regime, which is of course a first
requisite. We define the peculiar velocity ${\bf v}$ by: 
\beq 
\dot{\bf r} = {\bf u} = \dot{a} {\bf x} + a \dot{\bf x} =
\frac{\dot{a}}{a} {\bf r} + {\bf v} 
\eeq 
and the divergence $\theta$ ($\nabla = \partial/\partial {\bf x}$ in comoving coordinates, and $\nabla_r = \partial/\partial {\bf r}$) by: 
\beq 
\theta = \frac{\nabla {\bf v}}{\dot{a}} = \frac{a}{\dot{a}} \; \nabla_r {\bf v}
\eeq 
We can note that in the linear regime, where we keep only the growing
mode, we have: 
\beq 
\theta = - \delta \; \frac{a \dot{D}}{\dot{a} D} = - \delta \;
f(\Omega) 
\eeq 
where $D(t)$ is the growing mode of the density fluctuations and
$f(\Omega) \simeq \Omega^{0.6}$ (see Peebles 1980).

To use the method we described for the density contrast $\delta$, we
must now link the divergence $\theta$ of the velocity field to the
linear density contrast $\delta_L$ of our Lagrangian elements of
matter. One possibility is to make the approximation that the density
is uniform over these individual regions, and to use the continuity
equation (in physical coordinates): \beq \frac{\partial \rho}{\partial
t} + \rho \nabla_r {\bf u} = 0 \eeq An alternative is to consider the
mean divergence over the Lagrangian matter element $V$, which can be
expressed in terms of the expansion of this global volume (velocity of
the outer boundary): 
\beq 
{\dot{a}} \; < \theta > = \frac{1}{V} \int_V (\nabla . {\bf v}) \; d^3
x = \frac{1}{V} \int_S {\bf v} . {\bf ds} 
\eeq 
This last ``definition'' of $\theta$ is the most natural as it does
not need any additional information on the density profile. However,
in both cases we obtain: 
\beq 
\theta = - \frac{dln (1+\delta)}{dln \; a}
\label{theta} 
\eeq 
which relates $\theta$ to $\delta_L$ through the relation $\delta =
{\cal F}(\delta_L,a)$ used in the previous sections (note that the
derivative in (\ref{theta}) is to be understood at fixed $\nu$: one
follows the motion of a given fluid element, so that $\delta_L$ is
also a function of $a$). Then, we obtain the Lagrangian probability
distribution $P_m(\theta) \; d\theta = - P_L(\delta_L) \; d\delta_L$
and its Eulerian counterpart, which is simply given by $P(\theta) \;
d\theta = - P(\delta) \; d\delta$ (we introduced a negative sign
because $d\theta/d\delta <0$). Thus the latter can be written: 
\beq
P(\theta) = \frac{-1}{\sqrt{2\pi}} \; \frac{1}{1+\delta} \;
\frac{\partial}{\partial \theta} \left[ \frac{\delta_L}{\Sigma(R_m)}
\right] \; \exp\left[-\frac{1}{2} \left( \frac{\delta_L}{\Sigma}
\right)^2 \right] \label{Ptheta} 
\eeq 
We can also introduce the functions
$\varphi_{m\theta}(y,\xia_{m\theta})$ and
$\varphi_{\theta}(y,\xia_{\theta})$ (with $\xia_{\theta} =
<\theta^2>$). As was the case for the density contrast we recover in
the limit $\xia_{\theta} \rightarrow 0$ the functions derived by
Bernardeau (1994a).

If we make the approximation ${\cal F}(\delta_L,a) = {\cal
F}_0(\delta_L)$ where ${\cal F}_0$ is the function obtained in the
limit $\Omega \rightarrow 0$ (see (\ref{F0})) we get: 
\beq 
\theta = - f(\Omega) \; \delta_L \left(1-\frac{2}{3} \delta_L
\right)^{-1} 
\eeq 
Then, if we define $\varpi = \theta/f(\Omega)$ we obtain: 
\beq
P(\theta) \; d\theta = P(\varpi) \; d\varpi 
\eeq 
with
\[
P(\varpi) = \frac{1}{\sqrt{2\pi}\Sigma} \left(1-\frac{2}{3} \varpi
\right)^{(n-11)/4} \left(1-\frac{n+3}{6} \varpi \right)
\]
\beq
\hspace{2cm} \times \; \exp \left[ - \frac{\varpi^2}{2\Sigma^2}
\left(1-\frac{2}{3} \varpi \right)^{(n-1)/2} \right] \label{Pvarpi}
\eeq 
Note that the probability distribution of the reduced variable
$\varpi$ does not depend any longer on the cosmology (we only
neglected the slight dependence on $\Omega$ of the function ${\cal
F}$).

\begin{figure}[htb]

\begin{picture}(230,160)(-18,-19)

\epsfxsize=8.22 cm
\epsfysize=12 cm
\put(-14.5,-117){\epsfbox{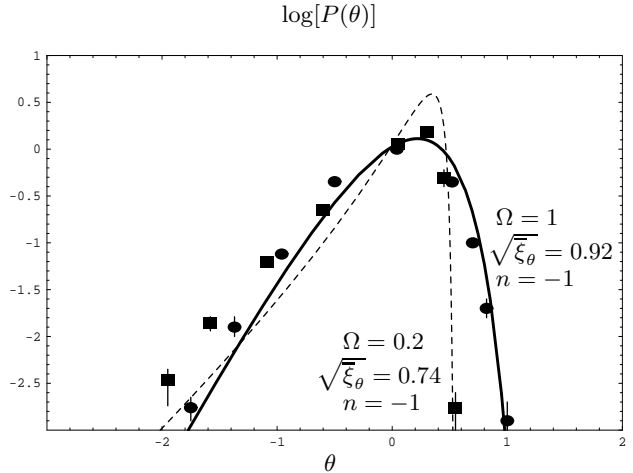}}
\put(89,140){$\log[P(\theta)]$}
\put(105,-30){$\theta$}
\put(170,63){$\Omega=1$}
\put(167,50){$\sqrt{\xia_{\theta}} = 0.92$}
\put(170,40){$n=-1$}
\put(112,16){$\Omega=0.2$}
\put(102,3){$\sqrt{\xia_{\theta}} = 0.74$}
\put(112,-7){$n=-1$}

\end{picture}

\caption{The probability distribution of the divergence of the
velocity field $P(\theta)$. The solid line is the prediction of our
prescription for a critical universe while the dashed-lines
corresponds to an open universe $\Omega=0.2 \; , \; \Lambda=0$. The
data points are taken from Bernardeau et al.(1997) for the same
conditions (filled circles for the critical universe and rectangles
for the open universe).}
\label{figPth1}

\end{figure}

We compare this approximation (\ref{Pvarpi}) to the results of
numerical simulations taken from Bernardeau et al.(1997) on
Fig.\ref{figPth1}. We can see that we match the high cutoff of
$P(\theta)$, which corresponds simply to the expansion rate of a void
(zero density), but the negative tail is not well reproduced for the case 
of the open universe. Thus, it seems that the
description of the velocity field is more sensitive on the
approximations involved in our method than the density field. This may
be due to the influence of the shear, which implies that the velocity
can no longer be determined by a mere scalar (through the potential or
the divergence). Moreover, we can note that, contrary to the Zeldovich
approximation for instance, our prescription does not provide the
location of particles and their velocity field from initial conditions
(all regions of space cannot follow a spherical dynamics at the same
time), it only gives an estimate for some probability distributions
without considering the consistent dynamics of all particles
simultaneously. Thus, the goal of our approach is more modest than
such a global modelization and it is not entirely
self-consistent. However, it appears that after accepting this
shortcomings we obtain nevertheless reasonable results for the density
field. In fact, as we shall see in the next section, we expect large
density fluctuations to follow a dynamics close to the spherical model
while the intermediate areas which connect these regions obey a
complex non-spherical dynamics but as their density contrast is of the
order of $\Sigma$ they do not play an important role for the global
shape of $P(\delta)$ as long as $\Sigma < 1$.

\section{Spherical density fluctuations}

As we showed in the previous paragraphs, the spherical dynamics
provides a good approximation to the behaviour of the density field 
(and a reasonable description for the velocity divergence) in
the quasi-linear regime, and it even gives the exact generating
functions $\varphi(y)$ in the limit $\xia \rightarrow 0$. This may
look surprising, since all regions of space cannot follow
simultaneously a spherical dynamics as we have already noticed, so
that one would expect this prescription to be always somewhat
different from the exact results. In fact, as we shall see below, this
is due to the fact that the limit $\xia \rightarrow 0$ (or
equivalently $\Sigma \rightarrow 0$) constrains the increasingly rare
density fluctuations of order unity to be spherically symmetric. Then
a spherical dynamics should naturally lead to correct results. Note that 
the mean evolution before virialization of rare large density fluctuations was
treated by Bernardeau 1994b.

Thus, let us define $\delta_L$ (resp. $\delta'_L$) as the mean density
contrast over a sphere of radius $R$ (resp. $R'$) centered on a point
O (resp. O'), and we note ${\bf r}$ the vector
$\underline{\mbox{OO'}}$. Then, the conditional probability
$P(\delta'_L|\delta_L)\;d\delta'_L$ to get a density contrast $\delta'_L$
knowing that we have a density contrast $\delta_L$ in the first sphere
is simply a gaussian: 
\beq 
P(\delta'_L|\delta_L) = \frac{1}{\sqrt{2\pi}\Sigma_0} \;
e^{-(\delta'_L-\delta_{L0})^2/(2\Sigma_0^2)}   \label{deltaL0}
\eeq 
where we defined:
\[
\Sigma^2=\int_0^{\infty} dk \; 4\pi k^2 \; W(kR)^2 \; P(k) \hspace{0.7cm} , 
\hspace{0.7cm} \Sigma'=\Sigma(R')
\]
\[
\Sigma''^2=\int_0^{\infty} dk \; 4\pi k^2 \; W(kR) W(kR') \; P(k) \;
\frac{\sin(kr)}{kr}
\]
\[
\mbox{hence} \;\; \Sigma = <\delta_L^2> \; , \; \Sigma' = <\delta_L'^2> \; , \; \Sigma'' = <\delta_L \; \delta_L'>
\]
\[
\mbox{and} \;\; \delta_{L0} = \delta_L \; \frac{\Sigma''^2}{\Sigma^2} \hspace{0.5cm} ,
 \hspace{0.5cm} \Sigma_0^2 = \frac{\Sigma^2 \Sigma'^2 -
\Sigma''^4}{\Sigma^2}
\]
and $W(x)=3(\sin x- x \cos x)/x^3$ is the top-hat window function (see for instance Bardeen et al.1986 for properties of gaussian random fields).
Thus, the mean value of $\delta'_L$ is $\delta_{L0}$ which only depends on
the distance $r=|{\bf r}|$ from the point O where the density contrast
is constrained to be $\delta_L$ over $R$, which is obvious from the
symmetry of the problem. However, the profile of the density
fluctuation centered on O is usually not spherically symmetric,
because of the fluctuations of $\delta'_L$. If we consider now the limit
$\Sigma \rightarrow 0$ (i.e. the normalization of the power-spectrum
goes to 0) at fixed $\delta_L$, the mean value $\delta_{L0}$ does not
change but $\Sigma_0 \rightarrow 0$. Thus, in this limit the profile
of the density fluctuation centered on O becomes spherically symmetric
($\delta'_L({\bf r}) = \delta_{L0}(|{\bf r}|)$), and its dynamics is
exactly described by the spherical model we used in the previous
paragraphs. Of course, most of the matter (and space) is formed by
density fluctuations $|\delta_L| \sim \Sigma$ which are not spherically
symmetric, but these areas correspond by definition to density
contrasts which tend to 0 as $\Sigma \rightarrow 0$ and they do not
influence the shape of the functions $\varphi(y)$ which depend on
finite values of the density contrast. This can be seen for instance
on (\ref{phiM}) which shows that a finite value of $y$ corresponds to
a finite value of the density contrast $\delta_y$ given by $\delta_y =
- y \; {\cal F}'(\delta_y)$ while $\Sigma \rightarrow 0$. Thus, our
approach explains the results of the rigorous calculation of the
generating functions $\varphi(y)$ in the limit $\Sigma \rightarrow 0$
from the exact equations of motion, and why they depend simply on the
spherical dynamics. Moreover, our model is not restricted a priori to
$\Sigma \ll 1$ and the comparison with numerical simulations shows it
gives reasonable results up to $\Sigma \sim 1$.

\section{Non-linear regime}

We showed in the previous paragraphs that a very simple model, based
on the spherical dynamics, can describe the evolution of the density
field up to $\xia \sim 1$. It is very tempting to try to extend this
model into the non-linear regime up to $\xia \sim 200$, which is
sufficient to describe approximately the subsequent highly non-linear
regime with the help of the stable-clustering ansatz (e.g.  
Peebles 1980 for a description of this latter approximation). However, this
implies that we take into account other processes like virialization,
which can only be made in a very crude way within the framework of the
prescription described in the previous paragraphs, so that we cannot
hope to get accurate quantitative results. In fact, this highly
non-linear regime where the probability distribution of the density
contrast is governed by the properties of virialized objects is
probably beyond the reach of rigorous perturbative methods based on
the expansion of the equations of motion given by a fluid description
with an irrotational velocity field. Indeed, the fluid approximation itself
breaks down after shell-crossing and cannot describe collapsing halos,
so that one has to use the Liouville equation which makes the analysis more difficult. Hence it may still be worthwile to consider
simple models like the one described in this article which could give
some hindsight into the relevant processes.

\subsection{Spherical collapse}

\subsubsection{Counts-in-cells}

According to the spherical dynamics large overdensities decouple
slowly from the general expansion of the universe, reach a maximum
radius $R_m$, turn-around and collapse to a singularity when their
linear density contrast $\delta_L$ is equal to a critical value
$\delta_c$, with $\delta_c \simeq 1.69$ for $\Omega=1$ (we shall only
consider the case of a critical universe in the following). However,
one usually assumes that such an overdensity will eventually virialize
(because the trajectories of particles are not purely radial) into a
finite radius to form a relaxed halo. Generally, this virialization
radius $R_v$ is taken to be one half of the turn-around radius, from
arguments based on the virial theorem, so we shall note $R_v =
\alpha/2 \; R_m$ with $\alpha \sim 1$. If we assume that after
virialization this halo remains stable, its density contrast will grow
as $a^3$, while its ``linear'' density contrast increases as
$a^2$. Thus, we modify the function $\delta={\cal F}(\delta_L)$ which
links $\delta_L$ to $\delta$ so that: 
\beq 
\delta_L \geq \delta_c \; : \;\; 1+\delta = \left(
\frac{10}{3\alpha} \right)^3 \; \delta_L^3 
\eeq 
At the time of collapse where the linear density contrast $\delta_L$
is equal to $\delta_c$, the actual density contrast of the halo is
$1+\delta = \alpha^{-3} \; (1+\Delta_c) \simeq \alpha^{-3}
178$. Within the framework of stable clustering, a similar relation
can be obtained for $\xia$ (see VS): 
\beq 
\xia > 200 \; : \;\; \xia(R) \simeq \left( \frac{10}{3\alpha_{\xi}}
\right)^3 \; \xia_L(R_L)^3 \label{xixiL} 
\eeq 
where $R_L$ is related to $R$ in a fashion similar to $R_m$ 
\beq 
R_L^3 = (1+\xia) \; R^3 
\eeq 
which comes from the fact that these approximations are based on a
Lagrangian point of view where one follows the evolution of matter
elements of constant mass. One can expect $\alpha_{\xi} \neq \alpha$
since all regions of space do not follow the same dynamics, and in
fact the previous model only applies to overdensities, which form one
half of the volume and mass in the early linear universe. Then, in a
fashion similar to what was done in VS, we obtain using (\ref{Pdel}):
\[
P(\delta) = \frac{1}{\xia^2} \; \frac{1}{\sqrt{2\pi}} \; \frac{5+n}{6}
\; \frac{\alpha}{\alpha_{\xi}} \; \left( \frac{\delta}{\xia}
\right)^{(n-7)/6}
\]
\beq
\hspace{2.5cm} \times \exp \left[ - \frac{\alpha^2}{2 \alpha_{\xi}^2}
\left( \frac{\delta}{\xia} \right)^{(5+n)/3} \right] \label{PdPS} 
\eeq
so that the probability distribution of the density contrast satisfies
the scaling-law: 
\beq 
P(\delta) = \frac{1}{\xia^{\;2}} \; h(x) \;\;\; \mbox{with} \;\;\;
x=\frac{1+\delta}{\xia} \label{Pdelhx} 
\eeq 
since in the regime we consider here ($\delta > 200, \xia >200$) we
have $(1+\delta)/\xia \simeq \delta/\xia$. The variable $x$ is related
to the usual linear parameter $\nu$ by: 
\beq 
\nu = \frac{\alpha}{\alpha_{\xi}} \; x^{(5+n)/6} 
\eeq 
while the scaling function $h(x)$ is simply: 
\beq 
x^2 h(x) = \frac{1}{\sqrt{2\pi}} \; \frac{5+n}{6} \; \frac{\alpha \;
x^{(5+n)/6}}{\alpha_{\xi}} \; \exp \left[- \frac{\alpha^2 \;
x^{(5+n)/3}}{2 \; \alpha_{\xi}^2} \right]
\label{hx} 
\eeq 
In fact, this scaling is characteristic of a much wider class of
models, defined by the scale-invariance of the many-body correlation
functions, studied in detail by Balian \& Schaeffer (1989). Thus, the
model we described in this article, based on the spherical dynamics
and a strong stable clustering assumption, appears as a simple way to
estimate the scaling function $h(x)$ characteristic of the highly
non-linear regime. Then, all the analysis developed for this general
class of density fields can be applied to this peculiar model. As was
noticed in VS, the approach presented in this article allows one to
recover the PS mass function as it is based on the same fundamental
idea: one follows the evolution of individual matter elements from the
linear regime, described by gaussian probability distributions, into
the non-linear regime. Indeed, if in a fashion similar to PS we identify 
the fraction $F(>M)$ of matter embedded within just-virialized objects of mass 
larger than $M$ with the mass contained in cells of scale $R$ with a density contrast $\delta > \Delta_{c \alpha}$ (with $(1+\Delta_{c \alpha}) = \alpha^{-3} \; \Delta_c \simeq \alpha^{-3} \; 178$ and $M=(1+\Delta_{c \alpha}) \rhoa V$) we obtain:
\beq
F(>M) = F(>\Delta_{c \alpha}, R) =  F_L(\delta_c,R_m)
\eeq
Here $F_L$ is the fraction of matter computed in the linear gaussian field, and we used the fact that in our spherical model a density contrast $\delta_L$ over a scale $R_m$ in the initial gaussian random field is associated to a density
contrast $\delta={\cal F}(\delta_L)$ over a scale $R$ in the actual non-linear density field as described in (\ref{FmFmLM}) and (\ref{deltLdelRmR}), as in the PS prescription. Then, the mass fraction in collapsed objects of mass $M$ to $M+dM$ is simply:
\beq
\begin{array}{ll}
{\displaystyle \mu(M) \frac{dM}{M} } &  {\displaystyle  = - \frac{\partial}{\partial R} F(>\Delta_{c \alpha}, R)  dR } \\ \\  &
{\displaystyle  = - \frac{\partial}{\partial R_m} F_L(>\delta_c, R_m)  dR_m }
\end{array}
\eeq
Since we have $F_L(>\delta_c, R_m) = P_L(>\delta_c, R_m)$ we recover as we should the PS mass function:
\beq
\mu(M) \frac{dM}{M} = \frac{1}{\sqrt{2\pi}} \; \frac{\delta_c}{\Sigma} \; \left| \frac{\mbox{dln}\Sigma}{\mbox{dln}M} \right| \; e^{-\delta_c^2/(2\Sigma^2)} \; \frac{dM}{M}
\eeq
However, (\ref{PdPS}) is not strictly equivalent to the 
usual PS prescription, since in addition to the spherical model it also 
relies on stable clustering. One should note that we did not
multiply our probability distribution by the usual factor 2, so that
we only recover one half of the total matter content of the universe
since the previous arguments only apply to initial
overdensities. Hence the scaling function (\ref{hx}) verifies $\int x
h(x) dx =1/2$ while this integral should be normalized to unity. This
implies (at least) that the approximate $h(x)$ we obtained cannot be
used for any $x$. In fact, following the discussion of Sect.4, we expect
the spherical dynamics model we used so far to be valid only for
extreme events $|\nu| \gg 1$. Hence the approximate $P(\delta)$ and
$h(x)$ we got should only apply to $x \gg 1$. The normalization
problem of the PS mass function is often ``cured'' by an overall
multiplication by a mere factor 2, ``justified'' by the excursion set
approach in the case of a top-hat in $k$ (Cole 1989; Bond et
al.1991). However, this result does not extend to other window
functions (like the top-hat in real space used here) with which for large
overdensities $\nu \gg 1$ the PS mass function does not suffer from
the cloud-in-cloud problem in the sense that the multiplicative factor 
needed to correct for double counting goes to 1 (and not 2) for large 
masses (Bond et al.1991; Peacock \&
Heavens 1990; VS). Hence, it appears that one should not multiply
(\ref{hx}) by 2, but merely restrict its application to $x \gg
1$. Moreover, we can note that this normalization problem is closely
related to the behaviour of underdensities, which are not well
described in the usual PS approach and constitute the missing half of
the matter content of the universe.

\subsubsection{Density profile of virialized halos}

We can notice that an overdensity with an initial density profile which is
exactly given by the mean value $\delta_{L0}$ (Sect.4) leads to a final
virialized halo with a flat slope in its inner parts, since the initial density
contrast $\delta_L'$ converges to a finite value at the center ($R' \rightarrow 0$ in the case $r=0$). This is not consistent with the results obtained from
numerical simulations which find inner density profiles $\rho \propto r^{-1}$
(Navarro et al.1996; Navarro et al.1997; Tormen et al.1997) or even steeper (Moore et al.1998). However, for moderate values of $\nu$ the correlation between the density at scale $R$ and the density at a smaller scale $R'$ 
enclosed within the former one quickly weakens as the fluctuation $\Sigma_0 \sim \Sigma'$ becomes larger than $\delta_{L0}$ ($\Sigma'$ diverges for small $R'$
while $\delta_{L0}$ remains finite). As a consequence one cannot infer the 
average density profile of virialized halos from $\delta_{L0}$. In fact, it seems more reasonable to consider an initial density profile of the form:
\beq
\delta_L \propto \Sigma \propto x^{-(n+3)/2}  \label{prof1}
\eeq
which leads to a final density profile for the virialized halo:
\beq
(1+\delta) \propto \rho \propto R^{-\gamma} \hspace{0.4cm} \mbox{with} 
\hspace{0.4cm} \gamma = \frac{3(3+n)}{5+n}   \label{slope}
\eeq
where $x$ (resp. $R$) is a comoving (resp. physical) coordinate. The reasoning
below (\ref{prof1}) is that if we look at a virialized halo of mass $M$, its characteristic density within a smaller sphere of mass $M'$ (centered on the 
peak) will be set by the maximum linear density contrast $\delta_L'$ realised
over all spheres of mass $M'$ enclosed within the larger matter element $M$.
The value of this maximum $\delta_{Lmax}'$ will scale with $M'$ as $\Sigma'$,
as soon as $\Sigma' \gg \Sigma$, see (\ref{deltaL0}), which leads to (\ref{prof1}). This picture also assumes that during virialization new collapsing shells which may not be centered on the density peak will roughly
circularize around it. We can note that for $-2<n<-1$ the slope (\ref{slope}) 
is $1< \gamma < 1.5$ which is close to what is seen in simulations (Navarro et al.1997; Moore et al.1998a). The same reasoning could also be applied to
underdensities, which we shall use in the next section. However, the previous arguments are quite crude
and a much more detailed analysis would be required to get a good description
of virialized halos. Moreover, within the approach described in the previous section the shape of the mean density profile has a rather weak meaning since
the model implies that a lot of substructure is present within halos, so
that a large object can be decomposed as a hierarchy of many smaller peaks 
with larger densities. Note that some simulations (Ghigna et al.1998) seem 
indeed to show that many sub-halos can survive within larger objects 
although not to such a large extent as in the model (this might be due to
finite resolution effects).

\subsection{Underdensities}

\subsubsection{Density profile of underdensities}

The function $x^2 \; h(x)$ obtained in the previous section (\ref{hx})
seems to show that the exponent of the small $x$ power-law tail is
$\omega=(5+n)/6$. However, as we argued above we do not expect this
scaling function to give reasonable results for $x \ll 1$, so that we
need to get $\omega$ from another point of view, which considers
explicitely low-density regions. Indeed, since we expect the spherical
dynamics approximation to be valid mainly for rare events we shall
shift from $\nu \gg 1$ to $\nu \ll -1$: that is we now study very
underdense areas. Moreover, initially non-spherical underdensities
tend to become increasingly spherical as they expand (contrary to the
collapse which enhances deviations from spherical symmetry), as seen
in Bertschinger (1985), so that a spherical dynamics model could
give reasonable results. We shall assume in this paragraph that the
many-body correlation functions are scale-invariant, so that the
density field is described by the scaling function $h(x)$ in the
non-linear regime $\xia \gg 1$ for $(1+\delta) \gg \xia^{\;
-\omega/(1-\omega)}$ which includes very underdense and small $x$
regions. Then, if we consider a sphere of radius $R$ with a density
contrast $\delta$ over $R$, the mean density contrast
$<1+\delta''>_{\delta}$ on its outer shell can be shown to be:
\beq 
x \ll 1 \; : \;\; <1+\delta''>_{\delta} = \left( 1+\frac{\gam}{3} \;
\frac{\omega}{1-\omega} \right) \; (1+\delta) \label{rhomega} 
\eeq 
for low density regions ($(1+\delta) \ll \xia$), where 
\beq
\gam = \frac{3 (3+n)}{5+n}
\eeq
is the slope of the two points correlation function in the highly
non-linear regime and $n$ is the slope of the initial power-spectrum
$P(k)$ which we assume here to be a power-law. This means that the
density profile is locally $\rho \propto R^{\;\gam
\omega/(1-\omega)}$. Thus, to get $\omega$ we may consider the
evolution of the density profile of a typical spherical underdensity
using (\ref{rhomega}). Let us follow an underdensity with an initial
profile in the early universe $|\delta_L| \propto \Sigma(M) \propto
M^{-(n+3)/6}$ (see (\ref{prof1})). Its dynamics is simply given
by the spherical model:
\beq 
\left\{ \begin{array}{lll} R & = A (\cosh \eta -1) \\ & & \;\;\;
\mbox{with} \;\; A^3={\cal G} MB^2 \\ t & = B(\sinh \eta -\eta)
\end{array} \right. \label{AA2} 
\eeq 
and 
\beq \delta_L(t) = -
\frac{3}{20} \left( \frac{6 t}{B} \right)^{2/3}
\label{AA4} 
\eeq 
Hence we have: 
\beq 
\left\{ \begin{array}{l} B \propto |\delta_L|^{-3/2} \\ \\ A \propto
M^{1/3} B^{2/3} \propto |\delta_L|^{-(n+5)/(n+3)} \end{array} \right.
\eeq 
At late times $\eta \gg 1$, and using (\ref{undersph}) we obtain: 
\beq
\left\{ \begin{array}{l} R \sim A \cosh \eta \sim A \; t/B \sim
|\delta_L|^{(n-1)/(2(3+n))} \\ \\ (1+\delta) \sim 1/\cosh \eta \sim B
\sim |\delta_L|^{-3/2} \end{array} \right.  \label{rhovoid} 
\eeq 
which means that $(1+\delta) \propto R^{3(3+n)/(1-n)}$. If we identify
this exponent with $\gam \omega/(1-\omega)$ we obtain eventually: 
\beq
\omega = \frac{5+n}{6} \label{omega} 
\eeq 
which is also the value one would have inferred from (\ref{hx}). We
can note that although the initial density is lower in the central
regions of the perturbation, which expand faster, there is no
shell-crossing and the density profile is given by (\ref{rhovoid}) for
$n<1$ (the profile of the initial density contrast is less steep than
$\delta_L \propto R^{-2}$). As we could expect from the analysis we
developed in the linear regime, we can note that the case $n=1$ leads
to some problems since we would have $\omega=1$. Hence it cannot be
described by this simple model. We shall only consider $-3<n<1$, which
corresponds to hierarchical clustering and cosmologically relevant
power-spectra (but one may expect that $n \rightarrow 1$ on very large
scales $R \rightarrow \infty$).

\subsubsection{Contact of underdensities}

We shall now develop a slightly more detailed description of the fate
of underdensities, in order to follow the behaviour of the matter (one
half of the total mass) which was initially embedded in these
low-density areas. This means that we have to modify the function
${\cal F}(\delta_L)$ for $\delta_L <0$ too. As we noticed in Sect.2,
underdensities grow very fast according to the spherical dynamics
which leads to an approximate probability distribution of the density
contrast with an ever increasing normalization (instead of unity). In
fact, the volume formed by any given range of underdensities will
eventually outgrow the available volume of the universe within the
approximation used so far. Indeed, we can write (\ref{Pnu1}) and
(\ref{Pdel}) as: 
\beq 
P(\delta) \; d\delta = \frac{1}{\sqrt{2\pi}} \; \frac{1}{1+\delta} \;
e^{-\nu^2/2} \; d\nu \label{Pnu2} 
\eeq 
with 
\beq 
\nu = \frac{\delta_L}{\Sigma(R_m)} = \frac{(1+\delta)^{(3+n)/6}
\delta_L} {\Sigma(R)} \label{nudelta} 
\eeq 
The linear variable $\nu$ defines the underdensities as it is a
constant of the dynamics. We can see from (\ref{Pnu2}) that a
logarithmic interval of $|\nu|$ of order unity will occupy all the
volume of the universe when 
\beq 
\frac{1}{\sqrt{2\pi}} \; \frac{|\nu|}{1+\delta} \; e^{-\nu^2/2} \sim 1
\eeq 
Naturally, we cannot keep our model unchanged for later times, since
the mean volume of these underdensities should not grow faster than
$a^3$ after this date. In fact, in this picture such a range of
negative density fluctuations first expands following the spherical
dynamics until neighbouring underdensities come into contact and fill
the entire universe. At this time the universe appears to be
constituted of low density bubbles of size $R$ and density $\rho$
(with $\rho < \rhoa$). The matter which was ``pushed'' by these
``voids'' to form the interface between adjacent underdensities gets
squeezed and reaches high densities. Thus, we shall assume the latter
build virialized structures of size $\sim R$ and density $\sim \rho$
(for instance they may have a density $\sim 200$ times higher than the
density $\rho$ of neighbouring areas). These represent the sheets and
filaments one can observe in numerical simulations (e.g. Bond et al.1996) 
which separate low-density regions, while the spherical overdensities
described in the previous sections are the nodes which form at their
intersections (note that the latter, corresponding to large positive initial density fluctuations, appear first). To see more clearly what this would imply for the
probability distribution of the density contrast (and for the mass
functions) we shall simply consider that underdensities defined by
their linear parameter $\nu$, and their scale $R$, stop expanding
when: 
\beq 
\frac{|\nu|}{1+\delta} \; e^{-\nu^2/2} = \beta
\label{nubeta} 
\eeq 
with $\beta \sim 1$ (that is when they fill all of the universe) and
keep after this date a constant radius $R$ and density $\rho$. Thus,
as time goes on the universe gets filled with increasingly large and
underdense ``bubbles'' while a growing fraction of the matter
progressively forms virialized structures (with a characteristic
density which becomes vanishingly small as compared to the mean
density). Note that within such a picture all the mass will eventually
be embedded in virialized high density objects, so that the usual
normalization problem is solved in a natural way. Thus, we must now
use (\ref{nubeta}) to obtain a relation $\nu - \delta$ in order to get
the probability distribution of the density contrast from
(\ref{Pnu2}). A negative density fluctuation $\nu$ ``stops'' when it
reaches a density contrast $\delta_s$ on a scale $R$ such that
$\Sigma=\Sigma_s$, defined by (\ref{nudelta}) and (\ref{nubeta}). At
late times, $\eta \gg 1$, the spherical dynamics (see
(\ref{undersph})) leads to the approximate relation: 
\beq 
(1+\delta) = \left( \frac{20}{27} |\nu| \Sigma(R) \right)^{6/(n-1)}
\eeq 
hence we obtain: 
\beq 
\left\{ \begin{array}{l} (1+\delta_s) = \frac{|\nu|}{\beta} \;
e^{-\nu^2/2} \\ \\ \Sigma_s = \frac{27}{20} \; \beta^{(1-n)/6} \;
|\nu|^{(n-7)/6} \; e^{(1-n) \; \nu^2/12} \end{array} \right.  
\eeq 
Note that the scale $R$ is not the Lagrangian scale $R_m$, and $\nu
\Sigma(R) \neq \delta_L$. After this ``stopping time'' $t_s$, the
radius of the object does not evolve any longer while its density
contrast increases as $a^3$, hence initial underdensities which have
already ``stopped'' verify: 
\beq 
(1+\delta) = (1+\delta_s)(\nu) \; \left(
\frac{\Sigma(R)}{\Sigma_s(\nu)} \right)^{6/(5+n)} 
\eeq 
since $\Sigma(R) \propto a^{(5+n)/2} \; R^{-(n+3)/2}$ by definition, where $R$ is the physical radius. This is the relation
$\nu - \delta$ we needed to derive the probability distribution of the
density contrast. We can note that the density contrast is now a
function of both $\delta_L$ and $\Sigma$, contrary to the pure
spherical dynamics case used previously where we had $\delta={\cal
F}(\delta_L)$. This is in fact a necessary condition to be able to get
eventually all the mass of the universe within overdense virialized
structures. Finally, this leads to: 
\beq 
P(\delta) \simeq \frac{1}{\xia^{\;2}} \; \frac{\omega}{\sqrt{2\pi}} \;
\frac{9\beta}{2\alpha_{\xi}} \; \left[ -2 \; \mbox{ln} \left(
\frac{9\beta}{2\alpha_{\xi}} x^{\omega} \right) \right]^{-3/2} \;
x^{\omega-2} 
\eeq 
where $\omega$ is defined by (\ref{omega}), $x$ by (\ref{Pdelhx}) and
we used (\ref{xixiL}) to introduce $\xia$. Thus, we recover the
scaling-law (\ref{Pdelhx}) with the same exponent $\omega$ as
previously, with logarithmic corrections. Such logarithmic terms may
indeed exist, but our model is probably too crude to give a reliable
estimate of their importance. As was the case for overdensities, we
also obtain a relation between the linear and non-linear parameters
$\nu$ and $x$: 
\beq 
\nu^2 \; e^{-\nu^2/2} = \frac{9\beta}{2\alpha_{\xi}} \; x^{\omega}
\label{nux} 
\eeq 
This scaling (\ref{Pdelhx}) of $P(\delta)$ applies to density
contrasts larger than the one of underdense regions which are
currently on the verge of filling the entire universe: $(1+\delta) >
(1+\delta_s)$ with: 
\beq 
(1+\delta) > (1+\delta_s) \sim (\mbox{ln}\xia)^{1/(2\omega-2)} \;
\xia^{\;-\omega/(1-\omega)} 
\eeq 
hence 
\beq 
x \gg \xia^{\;-1/(1-\omega)} 
\eeq 
which is exactly the limit predicted by a general study of the models
defined by the scale-invariance of the many-body correlation
functions. Thus, the approach developed in this section ``explains''
in a natural way both the emergence of the scaling-law (\ref{Pdelhx})
for small $x$ and its range of validity. Note that (\ref{hx}) derived
from the behaviour of overdensities only applied to virialized objects
$\delta > 200$, and was in fact restricted to $x \gg 1$ as we argued
above. The previous considerations also mean that, when seen on comoving 
scale $x$ at a time defined by the scale factor $a$, the universe appears to
be covered by very underdense regions of typical density contrast $\delta_s$ with (omitting logarithmic terms):
\beq
(1+\delta_s) \sim \Sigma(x,a)^{6/(n-1)} \sim a^{6/(n-1)} \; x^{3 (n+3)/(1-n)}
\label{deltas3D}
\eeq
as long as one remains in the non-linear regime ($\Sigma(x,a) \gg 1$ that is $(1+\delta_s) \ll 1$). The ``overdensity'' $(1+\delta_s)$ tends to 0 for large times or small scales as it should, while the volume occupied by most of the matter becomes increasingly negligible.

\subsubsection{Adhesion model}

We shall now try to compare the approach developed above to a very
different point of view: the adhesion model, in the peculiar case of
1-dimensional fluctuations (in a 3-dimensional universe). Indeed, from
the picture developed in the previous paragraphs we do not expect the
adhesion model to give reliable estimates for the counts-in-cells at
large values of $x > 1$, since in this range we should count roughly
spherical virialized halos which have a radius larger than the
considered scale $R$: these correspond to the overdensities described
by the spherical collapse seen in Sect.5.1. Hence the finite value of
the virialization radius of these objects plays a crucial role in the
final probability distribution of the density contrast, which is then
out of reach of the adhesion model where this radius is simply
zero. However, this model could give a fairly good picture of the
filaments and sheets which characterize the highly non-linear universe
as the transverse thickness of these structures, smaller than their
length or the radius of the neighbouring ``bubbles'', does not play an
important role on the counts-in-cells realised at these latter
scales. More precisely, we shall try to evaluate the typical density
contrast seen on a given scale: this corresponds to the maximum of
$P(\delta)$ and to the density contrast $\delta_s$ of the ``bubbles''
which cover the universe.

We shall first consider 1-dimensional density fluctuations, since in
this case the Zeldovich approximation is correct until shell-crossing
occurs so that we expect the adhesion model to provide reliable
results. We still define $n$ as the index of the power spectrum
$<|\delta_k|^2> \propto k^n$ so that we obtain: 
\beq 
-1<n<3 \; : \hspace{0.5cm} \Sigma(x) \propto a \; x^{-(n+1)/2}
\label{Sigad} 
\eeq 
This range in $n$ ensures that the density fluctuations increase at
small scales while the fluctuations of the density potential grow at
large scales, so that we are in the domain of the usual hierarchical
clustering. We note $x$ (resp. $q$) the comoving Eulerian (resp. initial
Lagrangian) coordinate of particles, in this 1-dimensional
problem (at early times $x \rightarrow q$). The relation $q-x$ is given by the adhesion model for the
displacement field, and the density is simply obtained from:
\beq
\eta(x,a) = \left| \frac{\partial q}{\partial x} \right| =  \frac{\partial q}{\partial x}
\hspace{0.5cm} \mbox{with} \hspace{0.5cm} \eta = (1+\delta)
\eeq 
We shall follow some of the notations used by Vergassola et
al.(1994) and we introduce the reduced velocity $v$ and velocity
potential $\psi$:
\beq
v(x,a) \equiv - \frac{\partial \psi}{\partial x} = \frac{V}{a {\dot a}}
\eeq
where $V=a {\dot x}$ is the peculiar velocity. We shall only consider
the cases $-1<n<1$ where the initial velocity ($a \rightarrow 0$) has
homogeneous increments and verifies the scale-invariance:
\beq 
\lambda>0 \; : \;\;  v_0(x+\lambda l) - v_0(x) \law
\lambda^{(1-n)/2} \; [ v_0(x+l)-v_0(x)] \label{v0} 
\eeq 
while the initial velocity potential satisfies: 
\beq
\lambda>0 \; : \hspace{0.5cm} \psi_0(\lambda x) \law \lambda^{(3-n)/2}
\; \psi_0(x) \label{psi0} 
\eeq 
with $v_0(0) = \psi_0(0)=0$. Here,
$\law$ means ``having the same statistical properties''. Then,
the overall density over the cell $[x_1,x_2]$ is simply: 
\beq
\eta([x_1,x_2],a) = \frac{1}{x_2-x_1} \; [ q(x_2,a)-q(x_1,a) ]
\eeq 
where the point $q(x,a)$ satisfies: 
\beq 
\left[ -\frac{q^2}{2} + a \psi_0(q) + x q \right] \; \mbox{is maximum at}
\; q(x,a) 
\eeq 
as given by the Hopf-Cole solution of the Burgers equation
(Hopf(1950), Cole(1951)) obtained from the adhesion model. If there is
a shock at the Eulerian location $x$, several Lagrangian coordinates
$q$ correspond to the same $x$ and we choose the smallest one which we
note $q_m(x,a)$. Then, we have:
\beq 
x = q_m + a \; v_0(q_m) \label{qmx} 
\eeq 
(which is simply the Zeldovich dynamics).  Using (\ref{psi0}) we can
check that the density $\eta(\Delta x,a)$ over a cell of size
$\Delta x$ satisfies:
\beq 
\eta(\Delta x,a) \law \eta(a^{-2/(n+1)} \; \Delta x \; , \; 1)
\label{etax} 
\eeq 
so that one only needs to consider the time $a=1$. The highly
non-linear regime which is of interest to us here corresponds to $a
\rightarrow \infty$ or $\Delta x \rightarrow 0$. Numerical simulations and
theoretical arguments
(She et al.1992; Vergassola et al.1994) strongly suggest that the Lagrangian
map $q \rightarrow x(q,1)$ forms a Devil's staircase and that shock
locations are dense in Eulerian space (for $-1<n<1$), which can be proved
rigorously for $n=0$ (Sinai 1992). Hence for almost every Eulerian
coordinate $x$ a Lagrangian coordinate $q_m$ exists. Using (\ref{qmx})
we can see that for two points $x_1$ and $x_2$ at time $a=1$ we have:
\beq 
x_2 - x_1 = q_{m2} - q_{m1} + [ v_0(q_{m2}) - v_0(q_{m1}) ]
\label{x2x1} 
\eeq 
In the limits $|x_2-x_1| \rightarrow 0$ and $|q_{m2}-q_{m1}| \rightarrow 0$
we have from (\ref{v0}) the behaviour
$|v_0(q_{m2})-v_0(q_{m1})| \sim |q_{m2}-q_{m1}|^{(1-n)/2}$. Since
$n>-1$ the second term in (\ref{x2x1}) becomes negligible, so that we
expect: 
\beq 
\Delta x \sim (\Delta q)^{(1-n)/2} \;\;\;  \mbox{and} \;\;\;  \eta^*(\Delta
x,1) \sim (\Delta x)^{2/(1-n)-1} 
\eeq 
where $\eta^*$ is the most probable value of $\eta$ (note however that
by definition the mean value of $\eta$ is simply $<\eta>=1$). Using
(\ref{etax}) we obtain: 
\beq 
(1+\delta)^*(\Delta x,a)  \sim
a^{-2/(1-n)} \; (\Delta x)^{(1+n)/(1-n)}
\label{del*} 
\eeq

We shall now consider the approach based on the spherical dynamics we
described in the previous sections, applied to this 1-dimensional
problem in order to compare its prediction to (\ref{del*}). The
equation of motion of a 1-dimensional density fluctuation, of
longitudinal size $2 R$, centered on the origin, in an universe
invariant by transverse translations, can be written: 
\beq 
\ddot{R} = \frac{4}{9} \; \frac{R}{t^2} \; - \; \frac{2}{3} \;
\frac{R_b}{t^2}
\label{Rmotion} 
\eeq 
where $R_b$ corresponds to the Hubble expansion of the matter element:
$R_b \propto a$ and $R/R_b \rightarrow 1$ as $t \rightarrow 0$ (on the
other hand the comoving transverse coordinates remain constant in
time). Thus, $x=R/R_b$ is the comoving coordinate of the outer front
which we normalized so that $x \rightarrow 1$ for $t \rightarrow
0$. We can write (\ref{Rmotion}) as: 
\beq 
3 \; t^2 \; \ddot{x} + 4 \; t \; \dot{x} - 2 \; x + 2 = 0
\label{xmotion} 
\eeq 
The solution of this equation is simply related to the initial
conditions by: 
\beq 
x = 1 + \left( \frac{t}{t_*} \right)^{2/3} \;\; \mbox{and} \;\;
\delta_L = - \left( \frac{t}{t_*} \right)^{2/3} 
\eeq 
where $\delta_L$ is the linear theory density contrast. Indeed, we
have $(1+\delta) = R_b/R = 1/x$. Thus, we obtain: $(1+\delta) =
(1-\delta_L)^{-1}$, so that at late times: 
\beq 
(1+\delta) \ll 1 \; : \;\; (1+\delta) \simeq |\delta_L|^{-1}
\label{deltadeltaL} 
\eeq 
The linear parameter $\nu$ is still given by $\nu =
\delta_L/\Sigma(R_m)$ where now $R_m = (1+\delta) R$ so that:
\beq
(1+\delta_s) \sim \left( |\nu| \Sigma(R) \right)^{2/(n-1)} 
\eeq 
Hence according to the model described in the previous section
underdensities stop expanding and fill the entire universe when they
reach the density contrast $\delta_s$ such that:
\beq 
(1+\delta_s) \sim a^{-2/(1-n)} \; x^{(1+n)/(1-n)}
\eeq 
where we omitted logarithmic corrections and we used
(\ref{Sigad}). Here, $x$ is the considered comoving scale we noted
$\Delta x$ in (\ref{del*}). Thus we recover exactly the behaviour seen
above within the framework of the adhesion model (\ref{del*}). This is
due to two effects: i) in such a 1-dimensional problem the Zeldovich
approximation gives the {\it exact} dynamics until a shock appears,
ii) the formation of these large underdensities corresponds to peaks
of the initial velocity potential which are global maxima over a large
scale but all particles located in the final broad low-density areas
come from a small Lagrangian region (the peak of $\psi_0$) so that
inner properties are given by {\it local} characteristics and the
spherical model is reasonable (the stop of the expansion at $\delta_s$
models the global constraints, related to the fact that these peaks
are only maxima over a finite scale). In fact, both models lead to
very similar pictures: most of the universe is filled by the expansion
of initial low-density peaks while most of the mass is squeezed
between this underdensities in virialized objects (for the spherical
model) or infinitesimally thin shocks (in the adhesion model).\\

For the usual 3-dimensional case, the adhesion model does not lead to
the same results as the spherical prescription and there would be a
break at $n=-1$ (while it occurs at $n=1$ for both models in the
1-dimensional case): indeed the previous arguments would give $\omega=(5+n)/4$
instead of (\ref{omega}) which leads to lower characteristic densities $(1+\delta_s)$. We think this discrepancy could be due to the
fact that the Zeldovich approximation does not give the exact dynamics
any more, even before a shock forms, so that the shape and size of
structures built at late times is not correctly described. Indeed, within this
approximation the physical coordinate ${\bf r}$ of a given particle evolves as:
\beq
{\bf r} = a ( {\bf q} + a {\bf p} )
\eeq
which means that for a spherical underdensity we obtain at late times 
$(1+\delta) \sim t^{-2}$ while the spherical dynamics leads to $(1+\delta) \sim t^{-1}$, see (\ref{rhovoid}). Hence the Zeldovich approximation overestimates
the expansion of voids which explains the low $(1+\delta_s)$ and the high 
$\omega$ it gives. Thus, the
adhesion model appears to confirm our spherical prescription for
1-dimensional fluctuations, where the former has a rather firm
foundation, which builds confidence in our model which predicts
moreover scaling-laws which are actually seen in numerical
simulations. For the generic 3-dimensional case, this latter result
suggests that our prescription is still valid, while the adhesion
model should worsen.

\subsubsection{Non-spherical corrections}

As we noticed above, we expect the spherical dynamics to describe
extreme events $|\nu| \gg 1$. Moreover, in the same way as the
``cloud-in-cloud'' problem is not very important for large
overdensities $\nu \gg 1$, as noticed in VS, which is necessary for
the approach developed in paragraph 5.1, the corresponding
``void-in-void'' problem disappears for $\nu \ll -1$ (since the
gaussian density field is symmetric under $\nu \leftrightarrow -\nu$)
which allows one to use the prescription presented in 5.2.2. However,
we can see from (\ref{nux}) that large negative values of $\nu$
correspond to very small $x$. Indeed, with $\omega \simeq 0.5$ we can
check that $\nu < -4$ leads to $x < 10^{-6}$, $(1+\delta_s) < 10^{-3}$
and $\xia > 10^3$. Thus, the range of $x$ which can be studied in
current numerical simulations $0.05 < x < 50$ may be too small to
recover the power-law tail with exponent $\omega$ and the exponential
cutoff predicted by our approach. Hence, non-spherical corrections may
change the value of $\omega$ obtained in numerical simulations. To get
an idea of the magnitude and direction of such effects, we can study
the case where underdensities only expand along 1 or 2 directions
(planar or cylindrical symmetry) while the other axis remain(s)
constant in comoving coordinates. Thus the 1-dimensional problem
considered in the previous section corresponds to the expansion along
only one direction, while the spherical model represents a growth
along all three axis.

For the 1-dimensional growth, (\ref{deltadeltaL}) implies that we have
$(1+\delta_s) \sim \Sigma(M)^{-1}$ where we do not
consider logarithmic corrections (i.e. $|\nu| \sim 1$). Using $R_m=R_b$ 
and $\Sigma(R_b) \propto \xia(R_b)^{(5+n)/6}$ we obtain $(1+\delta_s) \sim
\xia^{\;-(5+n)/6}$ on scale $R_b$ (the smallest scale of the
underdense regions). If we write this exponent as $-\omega/(1-\omega)$
we get: \beq 
\mbox{1-D} \; : \;\; \omega = \frac{5+n}{11+n} \label{1D}
\eeq 
In a similar fashion, a two-dimensional expansion leads to $(1+\delta)
\propto |\delta_L|^{-(\sqrt{13}-1)/2}$ and: 
\beq 
\mbox{2-D} \; : \;\; \omega = \frac{(5+n) (\sqrt{13}-1)}{(5+n)
(\sqrt{13}-1) + 12}
\label{2D} 
\eeq 
We compare on Table 3 the values obtained in numerical simulations to
these estimations. Thus, although we recover the increase of $\omega$
with $n$, non-spherical corrections appear to be non-negligible. Hence
we expect the value of $\omega$ measured in numerical simulations,
which is presently close to the 2-D result, to increase somewhat for
smaller $x$ at smaller scales where $\xia$ is larger, and to get closer to
the value (\ref{omega}) obtained from the spherical
dynamics model.

\begin{table}
\begin{center}
\caption{Exponent $\omega$ for various indexes $n$ of the
power-spectrum. The lines 1D, 2D and 3D corresponds to (\ref{1D}),
(\ref{2D}) and (\ref{omega}). The last two lines present the results
of numerical simulations: Colombi et al.(1997) for A and Munshi et
al.(1998) for B.}

\begin{tabular}{ccccc}\hline

$n$ & $-2$ & $-1$ & $0$ & $1$ \\ 
\hline\hline
\\ 

1D & 0.33 & 0.4 & 0.45 & 0.5 \\

2D & 0.39 & 0.46 & 0.52 & 0.62 \\

3D & 0.5 & 0.66 & 0.83 & 1  \\
\hline\hline
\\

A & 0.3 & 0.5 & 0.65 & 0.65 \\

B & 0.33 & 0.4 & 0.55 & 0.7

\end{tabular}
\end{center}
\label{table3}
\end{table}

\subsection{General picture}

Thus, in the non-linear regime virialized objects should form through
two different processes according to our model. First, large
overdensities with a roughly spherical shape collapse as in the PS
approach to build high-density virialized halos. This corresponds to
matter elements described by $\nu \gg 1$ and $x \gg 1$. Second, large
underdensities expand until they fill the entire universe and see
their dynamics influenced by their interaction with neigbhouring
``voids''. The matter ``pushed'' by these regions forms high-density
filaments and sheets at their interface, which builds virialized
structures of increasingly large scale and low density (increasingly
lower than the mean density of the universe). Obviously the dynamics
of the filaments and walls cannot be described by a spherical model,
but our approach takes advantage of the fact that the low-density
``bubbles'' they surround may still be described by a spherical
dynamics, with the addition of other processes to take into account
the global constraints which stop their expansion. This models matter
elements with $\nu \ll -1$ and $x \ll 1$. The behaviour of the
intermediate fluid elements $|\nu| \sim 1$ is certainly quite complex
and depends on non-local properties through the influence of
neigbhouring peaks and voids. It is not described by our model and
corresponds to the transition interval of the scaling function $h(x)$
around $x \sim 1$ from its exponential cutoff to its power-law
tail. Note that in the regime $\Sigma \gg 1$, low density contrast
regions $|\delta| \ll 1$ are not described by linear theory, even
though $|\delta|$ is small, because of shell-crossing. Thus, our
model does not provide as accurate predictions as the PS mass
function, but it can be tested in numerical simulations by studying
the density profiles of large halos or voids, as well as the
asymptotic behaviour of $h(x)$. It would also be of interest to follow
the evolution of initial extreme matter elements $|\nu| \gg
1$. Moreover, we think the explicit description of underdensities is a
necessary step, which was not considered in detail in the PS formulation. It
ensures in a natural way that all the mass will eventually be embedded
within high-density virialized structures (which occupy a negligible volume), 
so that there is no
normalization problem, and it allows to describe the complex structure
formed by low-density bubbles, filaments and walls, which is seen in
numerical simulations (Cole et al.1997; Weinberg et al.1996; Bond et al.1996). One can note that according to our model, substructures should exist
within large overdensities, which is not always seen in numerical
simulations. However, as Klypin et al.(1998) argue this may be due to
a lack of numerical resolution, moreover substructures do appear
through counts-in-cells numerical studies.\\

In the non-linear regime, we have only considered the case of a
critical universe so far. However, in a low-density universe when
$\Omega$ becomes small virialized structures no longer form as the
linear growth factor $D(t)$ tends to a constant: the density
perturbations freeze in comoving coordinate. Thus, on small scales
where the density fluctuations are large, structures formed early when
$\Omega \simeq 1$ so that the analysis developed above for a critical
universe can be applied. To get the characteristics of virialized
structures today at these scales one simply needs to consider these
early times and then extrapolate until today using the approximation
(which was used throughout above) that virialized structures keep a
constant scale and density while the mean density of the universe
decreases as $\rhoa \propto a^{-3}$. On larger scales, fluctuations
will never reach the non-linear regime since $\Sigma$ tends to a
constant as $t \rightarrow \infty$, which is nearly reached as soon as
$\Omega \sim 0.1$. Hence, on these scales one can simply use the
quasi-linear description developed in Sect.2.2. Thus, one gets a
complete picture of the density field in a low-density universe
(except for a transitory range where $\xia \sim 10$) within the
framework of the approximation developed in this article.

\section{Conclusion}

Thus, in this article we have developed a simple model for
hierarchical clustering based on a spherical dynamics. We have shown
it provides a reasonable approximation to the density field, and the
divergence of the velocity field, in the quasi-linear regime $\Sigma
<1$. Moreover, it allows one to recover the exact series of the
moments of the probability distribution of the density contrast in the
limit $\xia \rightarrow 0$ (contrary to other models like the
Zeldovich approximation for instance), and sheds some light on the
rigorous results. Then, we have developed a way to extend this model
into the non-linear regime, by taking into account the virialization
of high overdensities (as in the PS approach) and also the behaviour
of very underdense areas (in a way different from the PS
prescription). This implies a particular scaling in $x$ of the
counts-in-cells over a well-defined range which is indeed verified by
the results of numerical simulations. Thus, our model deals with the
evolution of the density field in both linear and non-linear regimes,
for any cosmological parameters (open or critical
universe) and provides a simple reference to which one could compare the
results of a more rigorous treatment. Naturally, our approximation should be tested in more
detail with numerical simulations. The density profile of very
underdense regions, the evolution with time of the scale and density
of typical ``voids'', the asymptotic behaviour of the scaling function
$h(x)$, should allow one to measure the influence of non-spherical
corrections and the effects of substructure disruption on the density
field, which are not well described by our model, and precise its
accuracy.

\end{document}